{}%disable hyper at arxiv
{}%use pdflatex at arxiv

\documentclass[11pt,a4paper]{article}
\newif\ifpublic\publictrue
\newif\ifworking\workingfalse
\usepackage{mathtools,mathrsfs,amsbsy,amssymb,latexsym,amsfonts,amscd}
\usepackage{amsmath}
\usepackage[nosort]{cite}
\usepackage{graphicx}
\usepackage{stackrel}
\usepackage{bm}
\usepackage[usenames,dvipsnames]{color}
\usepackage[normalem]{ulem}
\usepackage{pdfpages}
\usepackage{tikz}
\usetikzlibrary{arrows.meta}
\definecolor{linkcolor}{rgb}{0,0,0.6}
\usepackage[colorlinks=true,pdfstartview=FitV,linkcolor=linkcolor,
citecolor=linkcolor,urlcolor=linkcolor,hyperindex=true,hyperfigures=false]{hyperref}
\ifpublic\else\usepackage{showkeys}\fi

\usepackage{fancybox,fancyhdr}
\usepackage{esint}

\usepackage[font=small,labelfont=bf,width=0.85\textwidth]{caption}

\def\showkeysrefformat#1{{\normalfont\tiny\ttfamily#1}}
\makeatletter
\def\SK@@ref#1>#2\SK@{%
{\@inlabelfalse\leavevmode\vbox to\z@{%
\vss\SK@refcolor\rlap{\vrule\raise .75em%
\hbox{\showkeysrefformat{#2}}}}}}
\makeatother

\usepackage[a4paper,text={160mm,247mm},centering]{geometry}
\global\arraycolsep=2pt
\allowdisplaybreaks[3]

\numberwithin{equation}{section}

\expandafter\def\expandafter\bfseries\expandafter{\bfseries\ifmmode\else\boldmath\fi}
\expandafter\def\expandafter\mdseries\expandafter{\mdseries\ifmmode\else\unboldmath\fi}
\expandafter\def\expandafter\normalfont\expandafter{\normalfont\ifmmode\else\unboldmath\fi}

\def\beq{\begin{equation}}
\def\eeq{\end{equation}}

\def\beqz{\begin{equation*}}
\def\eeqz{\end{equation*}}

\def\bea{\begin{eqnarray}}
\def\eea{\end{eqnarray}}

\def\ha{\mbox{\small $\frac{1}{2}$}}

\def\id{\protect{{1 \kern-.28em {\rm l}}}}

\def\f{\mathfrak f}

\def\g{\mathfrak g}
\def\h{\mathfrak h}
\renewcommand{\Im}{\operatorname{Im}}

\newcommand{\Act}{\mathcal{S}}
\newcommand{\alg}[1]{\mathfrak{#1}}
\newcommand{\grp}[1]{\mathrm{#1}}

\DeclareMathOperator{\tr}{tr}
\DeclareMathOperator{\Tr}{Tr}
\DeclareMathOperator{\PP}{P}
\DeclareMathOperator{\ad}{ad}
\DeclareMathOperator{\Ker}{Ker}
\DeclareMathOperator{\Ima}{Im}
\DeclareMathOperator{\Order}{O}

\def\Ad{\textrm{Ad}}

\def\st{{{\cal T}}}
\newcommand{\ti}[1]{_{\bm{\underline{#1}}}}

\newcommand{\indrm}[1]{{\mathrm{\scriptscriptstyle{#1}}}}
\newcommand{\antidiag}{\operatorname{antidiag}}
\newcommand{\groot}{{\cal{H}}}
\newcommand{\alga}{{\mathsf{A}}}
\newcommand{\algb}{{\mathsf{B}}}
\ifpublic

\else

\newcounter{comcompt}
\newcounter{piq}
\newcounter{treqle}
\newcounter{boxq}
\fi

\usepackage{fancyhdr}
\ifpublic\else
\fi

\makeatletter
\let\@keywords\@empty
\let\@subject\@empty
\providecommand{\keywords}[1]{\gdef\@keywords{#1}}
\providecommand{\subject}[1]{\gdef\@subject{#1}}
\def\thetitle{\@title}
\def\theauthor{\@author}
\def\thesubject{\@subject}
\def\thedate{\@date}
\def\thekeywords{\@keywords}
\makeatother
\AtBeginDocument{
\hypersetup{pdftitle={\thetitle}}
\hypersetup{pdfauthor={\theauthor}}
\hypersetup{pdfsubject={\thesubject}}
\hypersetup{pdfkeywords={\thekeywords}}}

\ifworking
\newcommand{\work}[1]{\begingroup \makeatletter\def\f@size{8}
#1 \endgroup}
\else
\newcommand{\work}[1]{\ignorespaces}
\fi
\newcommand{\workstep}[1]{\scriptsize $\bullet$~#1 \makeatletter\def\f@size{8}
}
\title{Towards a quadratic Poisson algebra for the \texorpdfstring{\\}{}
subtracted classical monodromy of \texorpdfstring{\\}{}
Symmetric Space Sine-Gordon theories}

\begin{document}
\setcounter{tocdepth}{2}
\begin{center}

\vspace*{2cm}

\begingroup\Large\bfseries\thetitle\par\endgroup

\vspace{1.5cm}

\begingroup
F. Delduc$^*$\footnote{E-mail:~francois.delduc@ens-lyon.fr},
B. Hoare$^\dagger$\footnote{E-mail:~ben.hoare@durham.ac.uk},
M. Magro$^*$\footnote{E-mail:~marc.magro@ens-lyon.fr},

\endgroup

\vspace{1cm}

\begingroup
* \it ENS de Lyon, CNRS, Laboratoire de Physique, F-69342 Lyon, France
\\
${}^\dagger$ \it Department of Mathematical Sciences, Durham University,
Durham DH1 3LE, UK
\endgroup

\end{center}

\vspace{2cm}

\begin{abstract}
Symmetric Space Sine-Gordon theories are two-dimensional massive integrable field theories, generalising the Sine-Gordon and Complex Sine-Gordon theories.
To study their integrability properties on the real line, it is necessary to introduce a subtracted monodromy matrix.
Moreover, since the theories are not ultralocal, a regularisation is required to compute the Poisson algebra for the subtracted monodromy.
In this article, we regularise and compute this Poisson algebra for certain configurations, and show that it can both satisfy the Jacobi identity and imply the existence of an infinite number of conserved quantities in involution.
\end{abstract}

\setcounter{footnote}{0}

\newpage

\tableofcontents

\section{Introduction}

Symmetric Space Sine-Gordon (SSSG) theories are two-dimensional integrable field theories that generalise the Sine-Gordon and Complex Sine-Gordon theories.
At the Lagrangian level they are defined as $G/H$ gauged Wess-Zumino-Witten (WZW) models plus a potential term.
The Sine-Gordon and Complex Sine-Gordon theories correspond to $G= SO(2)$, $H=\emptyset$ and $G=SO(3)$, $H=SO(2)$ respectively.
In this article, we are interested in the classical integrability of these theories when the spatial coordinate $x$ takes values on the real line.
A peculiarity of these theories is that for certain solitonic configurations \cite{Hollowood:2009tw,Hollowood:2011fm,Hollowood:2013oca} the Lax matrix does not vanish as $x \to \pm \infty$.
In the Sine-Gordon case, the asymptotic value of the Lax matrix is not zero and is proportional to the mass parameter (see for instance \cite{fadtakbook87,Babebook}).
As we will recall, the situation for SSSG theories is the same up to a gauge transformation.
Indeed, the asymptotic value of the Lax matrix of SSSG theories is the sum of two terms \cite{Hollowood:2013oca}.
The first term is proportional to the mass and has the same structure as the asymptotic limit of the Sine-Gordon Lax matrix.
The second term is pure gauge and takes values in the Lie algebra $\h$ associated with the Lie group $H$.

As a consequence of this, the transition matrix, i.e., the ordered exponential of the Lax matrix between two points $x_-$ and $x_+$, is not well-defined as $x_\pm \to \pm \infty$.
In other words, the monodromy matrix does not exist when $x$ belongs to an infinite interval.
In such cases, we need to define a subtracted monodromy matrix (see, for instance, \cite{Faddeev:1979gh,Sklyanin:1980ij,fadtakbook87,Babebook}).
For the Sine-Gordon theory, this subtracted monodromy matrix is not conserved but its time evolution takes the form associated with the existence of a Lax pair.
Hence the eigenvalues of the subtracted monodromy matrix are conserved.
Furthermore, the form of the Poisson brackets of the subtracted monodromy matrix implies that these eigenvalues are in involution.

Our goal is to determine the Poisson brackets of the subtracted monodromy matrix of SSSG theories.
The results presented here extend those obtained in \cite{Hollowood:2010dt,Hollowood:2013oca}, in which a subtracted monodromy matrix was defined.
As will be explained in section~\ref{sec2}, the subtracted monodromy matrix that we work with differs from that constructed in \cite{Hollowood:2013oca}.
The freedom that exists in the definition of the subtracted monodromy matrix affects both its time evolution and its Poisson bracket.
However, as expected and as will be shown, this does not change the quantities that are both conserved and in involution.

\medskip

In order to compute the Poisson bracket, we will need to deal with the non-ultralocality of the SSSG theories.
This means \cite{Maillet:1985fn,Maillet:1985ek} that there is a term proportional to the derivative of the Dirac distribution in the Poisson bracket of the Lax matrix.
It follows that the Poisson bracket of a transition matrix between points $x_1$ and $x_2$ with a transition matrix between points $y_1$ and $y_2$ is ill-defined whenever one of the points $x_i$ coincides with one of the points $y_i$.
There exists a general prescription to regularise in order to define Poisson brackets of transition matrices with coinciding endpoints.
However, the Jacobi identity for the resulting bracket is not satisfied \cite{Maillet:1985fn,Maillet:1985ek}.

Within the non-ultralocal integrable field theories, SSSG theories belong to a special class.
In particular, it has been shown in \cite{Delduc:2012qb,Delduc:2012mk} that the non-ultralocality is mild.
Being mildly non-ultralocal means the Poisson bracket of the Lax matrix takes a form \cite{Delduc:2012qb,Delduc:2012mk} that allows the Poisson bracket of transition matrices with coinciding endpoints to be regularised while satisfying the Jacobi identity.
More precisely, the regularisation assumes that there exists a lattice description of the theory based on Freidel-Maillet quadratic algebras \cite{Freidel:1991jx,Freidel:1991jv}.
This regularisation makes use of a matrix $\alpha$, which is a split skew-symmetric solution of the modified classical Yang-Baxter (mCYBE) on the complexification of the Lie algebra $\g$ associated with the Lie group $G$.

Let us note that SSSG theories behave differently to integrable sigma models.
The former are massive theories already at the classical level and their non-ultralocality is mild.
The latter are massless at the classical level and, while their monodromy matrix on the infinite interval does exist, there is no known regularisation of the ill-defined Poisson brackets that satisfies Jacobi identity.

\medskip

The Sine-Gordon and Complex Sine-Gordon theories are special cases of SSSG theories.
It is therefore instructive to review some relevant facts about these theories.
The commonly used Lax matrix of the Sine-Gordon theory is ultralocal.
However, for the Sine-Gordon case, the SSSG Lax matrix is non-ultralocal.
These two Lax matrices are related by a formal gauge transformation \cite{Vicedo:2017cge}.

The first truly non-ultralocal SSSG theory studied in the literature is therefore the Complex Sine-Gordon theory.
Viewed as a SSSG theory, this theory is a $SO(3)/SO(2)$ gauged WZW model plus a potential term.
To recover the familiar action for a complex scalar field \cite{Pohlmeyer:1975nb,Lund:1976ze,Lund:1977dt,Getmanov:1977hk} we fix the $SO(2)$ gauge invariance and integrate out the non-dynamical gauge fields.
The results obtained in \cite{Maillet:1985ek} for this gauge-fixed action provide further motivation for the analysis in this article.
In particular, although the `standard' regularisation is used in \cite{Maillet:1985ek}, the Poisson bracket of the subtracted monodromy matrix constructed in \cite{Maillet:1985ek} satisfies the Jacobi identity.
A natural question is then whether this situation is a generic property of SSSG theories, or if it is special to the Complex Sine-Gordon theory.

\medskip

The plan of this article is the following.
We start in section \ref{sec2} by introducing the Lax matrix and defining a subtracted monodromy matrix for any SSSG theory.
Our discussion will mainly follow \cite{Hollowood:2013oca}, albeit with a minor modification.
In section \ref{sec3}, we briefly discuss the missing data required in order to fully compute the Poisson bracket of the monodromy and the relation to the correct definition \cite{Hollowood:2013oca} of the action.
To proceed, we restrict to particular configurations and recap the computation of the Poisson bracket of the Lax matrix.

In section \ref{sec4}, we use a `lattice' regularisation to compute the Poisson bracket of transition matrices on the lattice, and take the continuum limit.
Using this result, we take the infinite-interval limit in the section \ref{sec5}, arriving at the Poisson bracket of the subtracted monodromy matrix.
We analyse the form of this Poisson bracket in section \ref{sec6}, checking the Jacobi identity, and showing that it implies the existence of an infinite number of conserved quantities in involution when the theory is considered on an infinite interval.
We conclude with comments and an outlook in section \ref{secconc}.

\section{The subtracted monodromy}\label{sec2}

We start by constructing the subtracted monodromy matrix.
Our construction closely follows that in the reference \cite{Hollowood:2013oca} with minor differences.
We consider SSSG theories that are obtained as the Pohlmeyer reduction \cite{Pohlmeyer:1975nb,DAuria:1979ham,DAuria:1980iyh,Bakas:1995bm} (see also \cite{Fernandez-Pousa:1996aoa} and \cite{Miramontes:2008wt} for a review) of the symmetric space $\sigma$-model for a compact Riemannian symmetric space $F/G$.
The resulting model is a $G/H$ gauged WZW theory plus a potential term.

\subsection{Algebraic background}\label{sec:alg}

We denote by $\f$, $\g$, $\h$ the Lie algebras corresponding to the Lie groups $F$, $G$, $H$ respectively.
As usual, since $F/G$ is assumed to be a symmetric space, we have the orthogonal, with respect to the Killing form, direct sum decomposition
\beq\label{eq:d1}
\f =\f^{(0)} + \f^{(1)}
\eeq
where $\f^{(0)} = \g$.
The spaces $\f^{(0)}$ and $\f^{(1)}$ are eigenspaces of the involutive automorphism $\sigma$ with eigenvalues $1$ and $-1$ respectively.
We denote the corresponding projector onto $\alg{g}$ as $P^\alg{g}$.

The potential term is defined in terms of $\Omega$, a constant element of $\f^{(1)}$.
The adjoint action of $\Omega$ induces a second orthogonal decomposition of $\f$:
\beq \label{eq:d2}
\f = \f^{\perp} + \f^{\parallel}, \qquad \f^{\perp} = \Ker \, \ad_{\Omega},
\qquad \f^{\parallel} = \Ima \, \ad_{\Omega} ,
\eeq
with
\beq
[ \f^\perp, \f^\perp] \subset \f^\perp, \qquad [ \f^\perp, \f^\parallel] \subset \f^\parallel,
\eeq
that is $\f^\perp$ is a subalgebra of $\f$.

The symmetric space structure implies that the two decompositions~\eqref{eq:d1} and~\eqref{eq:d2} are compatible and we have that
\begin{equation}
\f = \g^\perp + \g^\parallel + \f^{(1)}{}^\perp + \f^{(1)}{}^\parallel,
\end{equation}
with
\begin{equation}
\begin{array}{c|cccc}
\quad [,] \subset \quad & \quad \g^\perp \quad & \quad \g^\parallel \quad & \quad \f^{(1)}{}^\perp \quad & \quad \f^{(1)}{}^\parallel \quad
\\\hline
\g^\perp & \g^\perp & \g^\parallel & \f^{(1)}{}^\perp & \f^{(1)}{}^\parallel
\\
\g^\parallel & \g^\parallel & \g & \f^{(1)}{}^\parallel & \f^{(1)}
\\
\f^{(1)}{}^\perp & \f^{(1)}{}^\perp & \f^{(1)}{}^\parallel & \g^\perp & \g^\parallel
\\
\f^{(1)}{}^\parallel & \f^{(1)}{}^\parallel & \f^{(1)} & \g^\parallel & \g
\end{array}
\end{equation}
The subgroup $H$ of $G$ is defined as
\beq
H = \{ h \in G \, | \, h \Omega h^{-1} = \Omega \}.
\eeq
Therefore $\h$ is the Lie subalgebra of $\g$ formed by the elements that commute with $\Omega$.
This implies that $\h = \g^\perp$.

To make certain computations more tractable, we make the additional assumption that $\f^{(1)}{}^\perp$ is one-dimensional, i.e., the symmetric space is rank 1, and we denote this abelian algebra with generator $\Omega$ as $\alg{a} = \f^{(1)}{}^\perp$.
In sections \ref{sec2} to \ref{sec4}, we will make no additional assumptions about the form of the symmetric space $F/G$.
From section \ref{sec5} onwards, we also assume that $\Omega$ can be normalised such that $[\Omega,[\Omega,\alga]] = -\alga$ for all $\alga \in \f^\parallel$, while we recall that $[\Omega,\alga] = 0$ for all $\alga \in \f^\perp$.
It follows that, in this case, the eigenvalues of the adjoint action $\ad_\Omega$ of $\Omega$ are 0 and $\pm i$.
Therefore, as a vector space, the complexified algebra admits the following decomposition
\beq\label{haompommd}
\f^{\mathbb{C}}=\h^{\mathbb{C}} + \alg{a}^{\mathbb{C}} + \omega^+ + \omega^-,
\eeq
where $\h^{\mathbb{C}}$ and $\alg{a}^\mathbb{C}$ are the complexifications of
$\h$ and $\alg{a}$ respectively, and $\omega^\pm = \Ker(\ad_\Omega \mp i )$.
Note that $\f^\parallel{}^\mathbb{C} = \omega^+ + \omega^-$, while $\f^\perp{}^\mathbb{C} = \h^\mathbb{C} + \alg{a}^\mathbb{C} = \Ker \ad_\Omega$.
We denote the corresponding projectors onto $\alg{h}$ and $\alg{a}$ as $P^{\alg{h}}$ and $P^{\alg{a}}$ respectively.
The commutation relations between elements of
these eigenspaces with the above assumptions are summarised
in the following table
\begin{equation}
\label{list com basis adomega}
\begin{array}{c|ccc}
\quad [,] \subset \quad & \quad \h \quad & \quad \omega^\pm \quad & \quad \alg{a} \quad
\\\hline
\h & \h & \omega^\pm & \{0\} %v2
\\
\omega^\pm & \omega^\pm & \{0\} & \omega^\pm %v2
\\
\omega^\mp & \omega^\mp & \h^\mathbb{C} + \alg{a}^\mathbb{C} & \omega^\mp
\\
\alg{a} & \{0\} & \omega^\pm & \{0\} %v2
\end{array}
\end{equation}

The spheres, with $F = SO(N+1)$, $G = SO(N)$, $H = SO(N-1)$, are examples of symmetric spaces that satisfy the properties outlined above, with explicit expressions given in appendix~\ref{spheres}.
The analysis presented here is expected to also generalise to other symmetric spaces with a careful treatment of the algebraic structure.

\subsection{Action, equations of motion and Lax connection}

The action of SSSG theories is given by
\unskip\footnote{For $k$ to be integer-quantized for simply-connected compact Lie groups we have that
\begin{equation*}
\tr\big(\alga\algb\big) = \frac{1}{2h^\vee} \Tr\big(\ad_\alga\ad_\algb\big) \qquad \alga,\algb \in \alg{g} ,
\end{equation*}
where $h^\vee$ is the dual Coxeter number of $\alg{g}$. \label{foot:kf}} \cite{Bakas:1995bm,Fernandez-Pousa:1996aoa} %v2
\begin{align}\nonumber
\Act_{\indrm{SSSG}} & = - \frac{k}{4\pi} \Big[\frac12 \int_\Sigma dt dx \, \tr \big(g^{-1}\partial_+ g g^{-1}\partial_- g\big)
+ \frac{1}{6} \int_B dt dx d\xi \, \epsilon^{ijk} \tr\big( g^{-1}\partial_i g [g^{-1}\partial_j g , g^{-1}\partial_k g] \big)
\\\label{action sssg}
& \qquad \qquad \qquad - \int_\Sigma dt dx \, \tr\big(A_- g^{-1} \partial_+ g - A_+ \partial_-gg^{-1} + A_- g^{-1} A_+ g - A_- A_+\big)
\\\nonumber
& \qquad \qquad \qquad + m^2 \int_\Sigma dt dx \, \tr(g^{-1} \Omega g \Omega) \Big] ,
\end{align}
where $m$ is a constant with dimension of mass, the group-valued field $g \in \grp{G}$ and the gauge field $A_\pm \in \alg{h}$.
The light-cone derivatives are given by $\partial_\pm = \partial_t \pm \partial_x$ and our conventions for Stokes' theorem are
\begin{equation}
\int_B dt dx d\xi \, \epsilon^{ijk} \partial_i B_{jk} = \int_\Sigma dt dx \, \epsilon^{\mu\nu} B_{\mu\nu} ,
\end{equation}
with $\epsilon^{tx} = +1$ ($\epsilon^{+-} = - \epsilon^{-+} = -\frac{1}{2}$).
The equations of motion for $g$ are
\begin{equation}
\partial_-(g^{-1}\partial_+ g + g^{-1} A_+ g) - \partial_+ A_-
+ [A_-,g^{-1}\partial_+ g + g^{-1} A_+ g] + m^2[g^{-1}\Omega g, \Omega] = 0 ,
\end{equation}
or equivalently
\begin{equation}
\partial_+(-\partial_- g g^{-1} + g A_- g^{-1}) - \partial_- A_+
+ [A_+,-\partial_- g g^{-1} + g A_- g^{-1}] - m^2[\Omega,g\Omega g^{-1}] = 0 ,
\end{equation}
while the equations of motion for $A_\pm$ are
\begin{equation} \label{eom of A}
A_+ = P^\alg{h}(g^{-1}\partial_+ g + g^{-1} A_+ g) , \qquad
A_- = P^\alg{h}(-\partial_- g g^{-1} + g A_- g^{-1}) .
\end{equation}
This action is invariant under the gauge transformations
\begin{equation} \label{gauge transfo}
g \to h^{-1} g h , \qquad A_\pm \to h^{-1} A_\pm h + h^{-1} \partial_\pm h , \qquad h(x,t) \in \grp{H}.
\end{equation}
We are interested in the case where the worldsheet $\Sigma$ is the plane.
For soliton solutions, the fields
have a non-trivial behaviour for $x \to \pm \infty$ (see \cite{Hollowood:1994vx,Hollowood:2009tw,Hollowood:2011fm,Hollowood:2013oca} for details).
This implies that the definition \eqref{action sssg} of the action has to be amended by certain `boundary' terms, which has an impact on the canonical analysis.
This will be briefly discussed in section \ref{sec3}.
The equations of motion in the bulk are not affected by this modification.

A Lax connection for these SSSG theories is
\begin{subequations}
\begin{align}
L_+(\lambda) &= g^{-1} \partial_+g + g^{-1}A_+g - \lambda m \Omega,\\
L_-(\lambda)&= A_- - \lambda^{-1} m g^{-1} \Omega g,
\end{align}
\end{subequations}
where $\lambda$ is the spectral parameter.
It is flat on-shell,
\beqz
\partial_+ L_- -\partial_- L_+ +[L_+,L_-]=0.
\eeqz
Writing $L_\pm = M \pm L$, the space and time components of the Lax connection are
\begin{subequations}
\begin{align}\label{laxmatrix}
L(\lambda) &= \ha( g^{-1} \partial_+ g + g^{-1}
A_+ g - A_- - \lambda m \Omega +\lambda^{-1} m \, g^{-1}
\Omega g),\\
M(\lambda) &= \ha( g^{-1} \partial_+ g + g^{-1}
A_+ g + A_- - \lambda m \Omega -\lambda^{-1} m \, g^{-1}
\Omega g).
\end{align}
\end{subequations}
Under the gauge transformation \eqref{gauge transfo}, the Lax connection transforms as
\beq
L_\pm \to h^{-1} L_\pm h + h^{-1} \partial_\pm h.
\eeq

As in \cite{Hollowood:2013oca} we consider fields whose asymptotic behaviour
as $x \to \pm \infty$ mimic configurations of minimal energy.
This means that as $x \to \pm \infty$, we require $g(x,t) \to H^\pm$ up to gauge transformations, where $H^\pm$ are constant elements of the Lie group $H$, and $A_\mu(x,t)$ to be pure gauge.
More precisely, we write
\begin{subequations}\label{asymptotic sssg}
\begin{eqnarray}
g (x,t) &\stackrel[x\to \pm\infty]{}{\simeq}& U^{\pm}(x,t) H^{\pm} (U^{\pm}(x,t))^{-1},\\
A_\mu(x,t) &\stackrel[x\to \pm\infty]{}{\simeq}& - \partial_\mu U^{\pm}(x,t) (U^{\pm}(x,t))^{-1},
\end{eqnarray}
\end{subequations}
with $U^\pm(x,t) \in H$.
Here $U^\pm(x,t)$ are defined in neighbourhoods of $\pm \infty$.

Using the fact that $\Omega$ commutes with elements of $H$, the asymptotic behaviour of the Lax connection is
\begin{subequations}
\label{lim gsg LMinfty}
\begin{eqnarray}
L(\lambda,x,t) &\stackrel[x\to \pm \infty]{}{\simeq} -\partial_x U^\pm(x,t) (U^\pm(x,t)) ^{-1}
- \ha m (\lambda - \lambda^{-1}) \Omega \equiv L^{\pm\infty}(\lambda,x,t),\\
M(\lambda,x,t) &\stackrel[x\to \pm \infty]{}{\simeq} -\partial_t U^\pm(x,t) (U^\pm(x,t))^{-1}
- \ha m (\lambda + \lambda^{-1}) \Omega \equiv M^{\pm\infty}(\lambda,x,t).
\label{Mpminfty gsg}
\end{eqnarray}
\end{subequations}

\subsection{Subtracted monodromy}

We now consider the differential equation
\beq
\partial_x {\Psi}(\lambda, x,t) = - L(\lambda,x,t){\Psi}(\lambda, x,t).
\eeq
The field ${\Psi}(\lambda, x,t)$ is the familiar extended solution for integrable field theories.
For $x_1 \leq x_2$, we have
\beq \label{transition matrix gsg}
\Psi(\lambda, x_2,t) = \overleftarrow{U}(x_1,x_2,-L(\lambda);t) \Psi(\lambda,x_1,t),
\eeq
where $ \overleftarrow{U}(x_1,x_2,-L(\lambda);t)$ is the ordered exponential of $-L$.
The transition matrix is then defined as
\beq \label{Tl gsg}
T_\ell(\lambda,t) \equiv \Psi(\lambda, +\ell,t) (\Psi(\lambda, -\ell,t))^{-1}
= \overleftarrow{U}(-\ell,+\ell,-L(\lambda);t).
\eeq
Since the Lax matrix \eqref{laxmatrix} does not vanish as $x \to \pm \infty$, the limit $\ell \to \infty$ of $T_\ell(\lambda,t)$ is not defined.
It is therefore impossible to define the monodromy on $\mathbb{R}$.
In such situations, we can remove the divergent part and define the subtracted monodromy (see,
for instance, \cite{Faddeev:1979gh,Sklyanin:1980ij,fadtakbook87,Babebook}).

To do this, we introduce the functions $\widehat{\Psi}^\pm(\lambda,x,t)$ and $\widehat{\Phi}^\pm(\lambda,x,t)$, which are defined in neighbourhoods of $\pm \infty$.
These functions are related as
\beq \label{psihat gsg}
\widehat{\Psi}^\pm(\lambda,x,t) \equiv U^\pm(x,t) \widehat{\Phi}^\pm(\lambda,x,t),
\eeq
so that the differential equations
\beq
\partial_x \widehat{\Psi}^{\pm}(\lambda, x,t) = - L^{\pm\infty}(\lambda,x,t)
\widehat{\Psi}^{\pm}(\lambda, x,t)
\eeq
are equivalent to
\beq\label{diffeqphi}
\partial_x \widehat{\Phi}^{\pm}(\lambda, x,t) = k(\lambda)
\Omega \widehat{\Phi}^{\pm}(\lambda, x,t),
\qquad
k(\lambda) = \ha m (\lambda-\lambda^{-1}).
\eeq
To prove this, we have used the fact that $U^\pm$ commute with $\Omega$.
The solutions of the differential equations~\eqref{diffeqphi} are
\beq \label{phihat gsg}
\widehat{\Phi}^{\pm}(\lambda, x,t) = \widehat{\Phi}^0(\lambda,x) \gamma^\pm(\lambda,t),
\qquad
\widehat{\Phi}^0(\lambda,x) \equiv \exp\bigl( k(\lambda) x \Omega \bigr).
\eeq
The `constants' of integration $\gamma^\pm(\lambda,t)$ are independent of $x$.
However, in principle, they can depend on $t$.
It follows that
\beq \label{Psihatpm gsg}
\widehat{\Psi}^\pm(\lambda,x,t) = U^\pm(x,t) \widehat{\Phi}^0(\lambda,x) \gamma^\pm(\lambda,t).
\eeq
These equalities should be understood in neighbourhoods of $\pm \infty$ as the functions $U^\pm$ are only defined on such neighbourhoods.

We then take $\ell$ to be large and introduce
\beq \label{eq228}
\st_\ell(\lambda,t) \equiv \gamma^+(\lambda,t) \bigl(\widehat{\Psi}^+(\lambda,+\ell,t)\bigr)^{-1}
T_\ell(\lambda,t) \widehat{\Psi}^-(\lambda,-\ell,t) (\gamma^-(\lambda,t))^{-1}.
\eeq
Note that we have defined $\st_\ell(\lambda,t)$ so that it does not depend on the `constants' of integration $\gamma^\pm(\lambda,t)$.
Indeed, taking into account \eqref{Tl gsg}, \eqref{psihat gsg} and \eqref{phihat gsg}, we have
\begin{subequations} \label{2eqresmono}
\beq \label{subT gsg 1}
\st_\ell(\lambda,t) = \big(
U^+(+\ell,t)\widehat{\Phi}^0(\lambda,+\ell) \big)^{-1}
T_\ell(\lambda,t)
U^-(-\ell,t)
\widehat{\Phi}^0(\lambda,-\ell).
\eeq
The subtracted monodromy on $\mathbb{R}$ is then defined as
\beq \label{final def resmono}
\st(\lambda,t) \equiv \lim\limits_{\ell \to \infty} \st_\ell(\lambda,t).
\eeq
\end{subequations}

\medskip

Let us note that the subtracted monodromy is gauge invariant.
Under the gauge transformation \eqref{gauge transfo} we have
\beq
U^\pm(x,t) \to (h^\pm(x,t))^{-1} U^\pm(x,t), \qquad H^\pm \to H^\pm,
\eeq
where $h^\pm(x,t)$ denote the asymptotic values of $h(x,t)$ as $x\to\pm\infty$.
Using \eqref{Psihatpm gsg}, we obtain the gauge transformation
\beq
\widehat{\Psi}^\pm(\lambda,x,t) \to (h^\pm(x,t))^{-1} \widehat{\Psi}^\pm(\lambda,x,t).
\eeq
On the other hand
\beq
\overleftarrow{U}(x_1,x_2,-L(\lambda);t) \to (h(x_2,t))^{-1}
\overleftarrow{U}(x_1,x_2,-L(\lambda);t) h(x_1,t).
\eeq
This means that
$T_\ell(\lambda,t) \to (h(+\ell,t))^{-1} T_\ell(\lambda,t) h(-\ell,t)$, hence $\st_\ell(\lambda,t) \to \st_\ell(\lambda,t)$ and
\begin{equation}
\st(\lambda,t) \to \st(\lambda,t).
\end{equation}
showing that the subtracted monodromy is gauge invariant.

\subsection{Time evolution and Lax pair}
\label{subsec: time evolution}

Recalling how $T_\ell$ defined in \eqref{Tl gsg} evolves with time
\beq \label{time derivative Tl}
\partial_t T_\ell(\lambda,t) = T_\ell(\lambda,t) M(\lambda, -\ell,t)
- M(\lambda, +\ell,t) T_\ell(\lambda,t),
\eeq
we find that the time evolution of $\st_\ell$ defined in \eqref{subT gsg 1} is
\beq \begin{split}
\partial_t \st_\ell(\lambda,t) & =
\st_\ell(\lambda,t) \Ad_{U^-(-\ell,t)\widehat{\Phi}^0(\lambda,-\ell)}^{-1}\big( M(\lambda, -\ell,t) + \partial_t U^-(-\ell,t) (U^-(-\ell,t))^{-1}\big)
\\ & \quad - \Ad_{U^+(+\ell,t)\widehat{\Phi}^0(\lambda,+\ell)}^{-1}\big( M(\lambda, +\ell,t) + \partial_t U^+(+\ell,t) (U^+(+\ell,t))^{-1}\big)\st_\ell(\lambda,t).
\end{split}\eeq
Substituting the asymptotic behaviour of the Lax connection~\eqref{lim gsg LMinfty} and using that $U^\pm$ and $\widehat{\Phi}^0$ commute with $\Omega$, we have that
\begin{equation}\begin{aligned}
\partial_t \st_\ell(\lambda,t) &\stackrel[\ell\to \infty]{}{\simeq}
- \widetilde{k}(\lambda) \st_\ell(\lambda,t)
\Omega
+ \widetilde{k}(\lambda) \Omega
\st_\ell(\lambda,t),
\qquad
\widetilde{k}(\lambda) = \ha m (\lambda +\lambda^{-1}),
\end{aligned}\end{equation}
hence
\beq \label{Lax equation gsg res mono}
\partial_t \st(\lambda,t) = [\widetilde{k}(\lambda)\Omega, \st(\lambda,t)
].
\eeq
This means that there exists a Lax pair $(\st(\lambda,t),\widetilde{k}(\lambda)\Omega)$ and thus an infinite number of conserved quantities which can be extracted from $\tr\st(\lambda)$.

\medskip

There is some freedom in the definition of the subtracted monodromy.
This freedom is associated with the `constants of integration' $\gamma^{\pm}(\lambda,t)$ appearing in \eqref{phihat gsg} (see also \eqref{eq228}).
We shall give two useful choices that exploit this freedom.

For the first, we define
\begin{equation}\label{stgamma}
\st_\gamma (\lambda,t) = (\gamma^+(\lambda))^{-1} \st(\lambda,t) \gamma^-(\lambda) ,
\end{equation}
with $\gamma^\pm(\lambda)$ independent of $t$ and commuting with $\Omega$.
The time evolution is left unchanged
\beq \label{Lax equation gsg res mono gamma}
\partial_t \st_\gamma(\lambda,t) = [\widetilde{k}(\lambda)\Omega, \st_\gamma(\lambda,t)
].
\eeq
Therefore, while $\tr \st_\gamma(\lambda,t) \neq \tr \st(\lambda,t)$, we also have that $\tr \st_\gamma(\lambda)$ gives us an infinite number of conserved quantities.
In sections \ref{sec3} to \ref{sec5}, we will work with the subtracted monodromy $\st(\lambda,t)$ and compute its Poisson bracket.
In section \ref{sec6}, we will then show that there exists a suitable choice of $\gamma^\pm(\lambda)$ such that the
conserved quantities $\tr\st_\gamma (\lambda,t)$ are in involution.

For the second, we define
\beq \label{tauomega}
\st_\Omega(\lambda,t) = e^{-t \widetilde{k}(\lambda)\Omega} \st(\lambda,t)
e^{t \widetilde{k}(\lambda)\Omega}.
\eeq
In contrast to the previous case, a consequence of \eqref{Lax equation gsg res mono} is that the quantity $\st_\Omega(\lambda,t)$ is conserved.
It is also immediate that $\tr\st(\lambda,t) = \tr\st_\Omega(\lambda,t)$.
The conserved subtracted monodromy matrix $\st_\Omega(\lambda,t)$ can be identified with the one defined in \cite{Hollowood:2013oca}.

Up to this point we have been working in the Lagrangian formalism, where the time evolution is governed by the variational equations of the action.
In sections \ref{sec3} and \ref{sec4} we will switch to the Hamiltonian formalism and proceed to compute the equal-time Poisson bracket of the subtracted monodromy matrix $\st(\lambda)$ with itself.
In subsection~\ref{sec:conserved}, we will also see that $\st_\Omega(\lambda,t)$ has the same equal-time Poisson bracket as $\st(\lambda)$, hence can be used to define the same set of conserved quantities in involution.
However, it should be noted that, as usual, the time evolution of $\st_\Omega(\lambda,t)$ in the Hamiltonian formalism will not simply be given by the Poisson bracket with the Hamiltonian due to the explicit dependence on $t$. %v2

\section{Poisson bracket of the Lax matrix}\label{sec3}

Our goal is to compute the Poisson brackets of the subtracted monodromy matrix.
An important property is that its definition depends on the quantities $U^\pm$, which describe the asymptotic behaviours of the fields through equation \eqref{asymptotic sssg}.
The presence of these non-trivial `boundary conditions' imply that the bulk action \eqref{action sssg} needs to be amended \cite{Hollowood:2013oca} (see also \cite{Hollowood:2011fm,Hollowood:2010dt}).
In particular, this is necessary in order to define the Wess-Zumino term \cite{Alekseev:1998mc}.
On the other hand, one can deduce from the study of boundary Wess-Zumino-Witten models that the modification of the action implies that the `bulk' symplectic structure also needs to be modified (see for instance \cite{Gawedzki:2001rm,Gawedzki:2001ye,Elitzur:2001qd,Figueroa-OFarrill:2005vws}).
This implies that the Poisson brackets of $U^\pm$ with, for instance, the Lax matrix cannot be determined from the canonical analysis of the action \eqref{action sssg}.

In this article, as a first step towards the full result, we shall carry out the computation for $U^\pm = \id$.
This means that asymptotically
\beq \label{eq31asval}
g (x,t) \stackrel[x\to \pm\infty]{}{\simeq} H^{\pm} \qquad \mbox{and}
\qquad
A_\mu(x,t) \stackrel[x\to \pm\infty]{}{\simeq} 0.
\eeq
The advantage of this simplification is that
no modification of the action \eqref{action sssg} is needed.
It follows that the subtracted monodromy is given by the limit of
\beq \label{subT gsg 2}
\st_\ell(\lambda,t) = (\widehat{\Phi}^0(\lambda,+\ell))^{-1}
T_\ell(\lambda,t)
\widehat{\Phi}^0(\lambda,-\ell)
\eeq
when $\ell \to \infty$.

\subsection{Canonical analysis}
The canonical analysis associated with the action \eqref{action sssg} has already been carried out in
\cite{Delduc:2012qb,Delduc:2012mk}.
For completeness, we review the main steps.

Let $\{T^{\hat{a}}\}$ be a basis of the Lie algebra $\g$, which we complement in order to form a basis $\{T^a\}$ of $\f$.
We make use of tensorial notation and define the quadratic Casimir of $\alg{f}$
\beq
C\ti{12} = \kappa_{ab} T^a \otimes T^b,
\qquad \kappa^{ab} = -\tr(T^a T^b),
\qquad \kappa^{ab} \kappa_{bc} = \delta^a_c.
\eeq
Recalling that matrices $O\ti{12} \in \alg{f}\otimes \alg{f}$ can be understood as the kernels of operators $O: \alg{f} \to \alg{f}$, we have
\begin{equation}
O \alga = - \tr\ti{2} (O\ti{12}(1\otimes \alga)) , \qquad O\ti{12} = O\ti{1}C\ti{12},
\end{equation}
so that the operator corresponding to the quadratic Casimir of $\alg{f}$ is the identity operator.
The quadratic Casimir of the subalgebra $\g$ of $\f$ is $C^{(00)}\ti{12} = \kappa_{\hat{a}\hat{b}} T^{\hat{a}} \otimes T^{\hat{b}}$, and the corresponding operator is the projector $P^{\alg{g}}$ introduced in subsection~\ref{sec:alg}.
We also define $C^{(11)}\ti{12} = C\ti{12} - C^{(00)}\ti{12}$, which belongs to $\f^{(1)}\otimes \f^{(1)}$,
\unskip\footnote{Since $\tr$ is proportional to the Killing form, see footnote \ref{foot:kf}, it follows from the structure of the symmetric space~\eqref{eq:d1} that $\tr(\f^{(0)}\f^{(1)}) = 0$, hence $C\ti{12} \in \f^{(0)}\otimes \f^{(0)} + \f^{(1)}\otimes \f^{(1)}$.} %v2
where $\f^{(1)}$ is introduced in equation~\eqref{eq:d1}, and whose corresponding operator is the projector onto $\alg{f}^{(1)}$.
The quadratic Casimir of the Lie algebra $\h$ is the kernel of $P^{\alg{h}}$, that is $C^\h\ti{12} = P^\h\ti{1}C\ti{12}$, and we also define the kernel of the projector onto the abelian algebra $P^{\alg{a}}$ to be $C^{\alg{a}}\ti{12} = P^{\alg{a}}\ti{1}C\ti{12}$.
We will use the property
\beq \label{property Casimir}
[C\ti{12}, \alga\ti{1} + \alga\ti{2}]=0,
\eeq
for any $\alga \in \f$, and similar identities for $C^{(00)}\ti{12}$ or $C^\h\ti{12}$ with $\alga\in \g$ or $\alga\in \h$, extensively.

Note that, except for the special cases introduced in the previous paragraph, we use the same symbol for an operator $O$ and its kernel $O\ti{12}$.

\paragraph{Phase space description.}

Let $\phi_i$ with $i\in \{1,\cdots, \mbox{dim} \, G\}$ be local coordinates on the Lie group $G$, $\partial^i= \frac{\partial}{\partial \phi_i}$ and $\pi^i(x)$ the momentum conjugate to the field $\phi_i(x)$ such that
\beq
\{ \pi^i(x), \phi_j(y) \} =\delta^i_j \delta_{xy}
\eeq
where $\delta_{xy} =\delta(x-y)$ is the Dirac distribution.
As usual, we introduce the $\g$-valued field
\beq
X = L_i^{\, a} \pi^i T_a, \qquad T_a = \kappa_{ab} T^b,
\eeq
where
\beq
g^{-1} \partial^i g = L^i_a T^a, \qquad
L^i_a L^a_j = \delta^i_j,
\qquad
L^i_a L_i^b =\delta^b_a.
\eeq
Following \cite{Delduc:2019bcl}, to work with the Wess-Zumino term, we introduce
the $\g$-valued quantity $W(x)$ through
\beq
\frac{1}{6} \int dt dx d\xi \, \epsilon^{ijk} \tr\big( g^{-1}\partial_i g [g^{-1}\partial_j g , g^{-1}\partial_k g] \big) = \int dt dx \, \tr(W, g^{-1} \partial_t g).
\eeq
The fields $g$, $X$ and $W$ have the following Poisson brackets (see, for instance, \cite{Delduc:2019bcl}):
\begin{subequations}
\begin{align}
\{ X\ti{1}(x), g\ti{2}(y) \} &= g\ti{2}(x) C^{(00)}\ti{12} \delta_{xy},\\
\{ X\ti{1}(x), X\ti{2}(y)\} &= - [C^{(00)}\ti{12}, X\ti{2}(x)] \delta_{xy},\\
\{ X\ti{1}(x), (g^{-1}\partial_x g)\ti{2}(x) \} &= - [C^{(00)}\ti{12} , (g^{-1}\partial_x g)\ti{2}(x)]
\delta_{xy} - C^{(00)}\ti{12} \partial_x \delta_{xy},\\
\{ X\ti{1}(x), W\ti{2}(y)\} + \{ W\ti{1}(x), X\ti{2}(y)\} &= - [C^{(00)}\ti{12},
W\ti{2}(x) - (g^{-1}\partial_x g)\ti{2}(x)] \delta_{xy}.
\end{align}
\end{subequations}

Introducing the rescaled level
\beq
K \equiv \frac{k}{4\pi},
\eeq
the currents
\begin{subequations}
\begin{align}
{\cal J}_L &= X + K W + K g^{-1} \partial_x g,\\
{\cal J}_R &= 2 K \partial_x g g^{-1} - g {\cal J}_L g^{-1},
\end{align}
\end{subequations}
form a pair of
commuting Kac-Moody currents with opposite level
\begin{subequations}
\begin{align}
\{ {{\cal J}_L}\ti{1}(x), {{\cal J}_L}\ti{2}(y) \}&=
- [C\ti{12}^{(00)} , {{\cal J}_L}\ti{2}(x) ] \delta(x-y) - 2 K C\ti{12}^{(00)} \partial_x \delta(x-y),\\
\{ {{\cal J}_R}\ti{1}(x), {{\cal J}_R}\ti{2}(y) \}&=
- [C\ti{12}^{(00)} , {{\cal J}_R}\ti{2}(x) ] \delta(x-y) + 2 K C\ti{12}^{(00)} \partial_x \delta(x-y),\\
\{ {{\cal J}_L}\ti{1}(x), {{\cal J}_R}\ti{2}(y) \}&= 0.
\end{align}
\end{subequations}

\paragraph{Hamiltonian.}
The relation between $X$ and $g^{-1} \partial_t g$ is
\beq \label{relgtimeder}
X= K( g^{-1} \partial_t g - W - A_- + g^{-1} A_+ g) .
\eeq
This identity encodes the relation
between $\pi^i(x)$ and $\dot{\phi}_j(x)$.
Therefore, the Lagrangian expressions for the currents ${\cal J}_L$ and ${\cal J}_R$ are
\begin{subequations}
\begin{align}
{\cal J}_L &= K(g^{-1} \partial_+ g - A_- + g^{-1} A_+g),\\
{\cal J}_R &= K(-\partial_- g g^{-1} + g A_- g^{-1} -A_+).
\end{align}
\end{subequations}

Letting $P_\pm$ denote the $\h$-valued fields `conjugate' to $A_\mp$ such that
\beq
\{ {P_\pm}\ti{1}(x) , {A_\mp}\ti{2}(y) \} = C\ti{12}^\h \delta_{xy},
\eeq
there are two primary constraints $\chi_1 \equiv P_+ \approx 0$ and
$\chi_2 \equiv P_- \approx 0$.
Imposing the stability of these constraints under time evolution leads to two secondary constraints $\chi_3 \approx 0$ and $\chi_4 \approx 0$ respectively with
\beq
\chi_3 \equiv P^\h {\cal J}_L - K(A_+-A_-) \qquad \mbox{and} \qquad
\chi_4 \equiv P^\h {\cal J}_R+K (A_+-A_-).
\eeq
One then introduces four Lagrange multipliers $\nu_i$ with $i=1,2,3,4$ and adds the sum $\sum_{i=1}^4 \nu_i \chi_i$ to the Hamiltonian density obtained by performing the Legendre transform of the Lagrangian.
The last step consists of studying the stability of all constraints $\chi_i$ under the time evolution generated by this Hamiltonian.
The two constraints $\chi_1$ and $\chi_2$ are stable provided that $\nu_3 = \nu_4$.
The two constraints $\chi_3$ and $\chi_4$ are stable provided that
\beqz
\nu_1 - \nu_2 = -\partial_x( 2 \nu_3 + A_+
+A_-) + [\nu_3, A_+ - A_-] -[A_+,A_-].
\eeqz
The Hamiltonian density is then found to be
\begin{equation}\begin{aligned}
H=& -\tr\Bigl(
\frac{1}{4 K} ({\cal J}_L^2 + {\cal J}_R^2)
+ {\cal J}_L A_- + {\cal J}_R A_+
+ \frac{K}{2} (A_+-A_-)^2 - K m^2 g^{-1} \Omega g \Omega \\
&\qquad + \nu_1 \chi_1 + \nu_2 \chi_2 + \nu_3 (\chi_3
+\chi_4)
\Bigr).
\end{aligned}\end{equation}
The Hamiltonian expression of the Lax matrix \eqref{laxmatrix} is then
\beq\label{eq:laxham}
L(\lambda) = \frac{1}{2K} {\cal J}_L - \ha \lambda m \Omega + \ha \lambda^{-1}
m g^{-1} \Omega g.
\eeq

\subsection{The \texorpdfstring{$r/s$}{r/s} structure}

We now have everything needed to compute the Poisson brackets of $L$.
The result can be put in the $r/s$ form of \cite{Maillet:1985fn,Maillet:1985ek}, i.e.
\begin{equation}\begin{aligned}\label{cont limit lattice rs}
\{ L\ti{1}(\lambda,x), L\ti{2}(\mu,y) \}
=\,& [r\ti{12}(\lambda,\mu), L\ti{1}(\lambda,x) + L\ti{2}(\mu,y)] \delta_{xy}
\\ & + [s\ti{12} , L\ti{1}(\lambda,x) - L\ti{2}(\mu,y)] \delta_{xy}
- 2 s\ti{12} \delta'_{xy},
\end{aligned}\end{equation}
with
\unskip\footnote{Note that for the Sine-Gordon model we have $F = SO(3)$, $G = SO(2)$, $H=\emptyset$.
In particular, $C\ti{12}^{(00)}$ does not vanish and the Poisson bracket for the Lax matrix is not ultralocal.
This Lax matrix is related to the familiar one, with ultralocal Poisson bracket, by a formal gauge transformation \cite{Vicedo:2017cge}.}
\beq \label{defofr12}
r\ti{12}(\lambda,\mu) =
\frac{1}{4K}\bigl( \frac{\mu^2+\lambda^2}{\mu^2-\lambda^2}
C\ti{12}^{(00)} + \frac{2 \lambda \mu}{\mu^2 -\lambda^2} C\ti{12}^{(11)}\bigr),
\qquad s\ti{12}= \frac{1}{4K}C\ti{12}^{(00)}.
\eeq
Indeed, the explicit computation of the Poisson bracket gives
\begin{equation}\begin{aligned} \label{pbLLcheck}
\{ L\ti{1}(\lambda,x), L\ti{2}(\mu,y) \}
= &- \frac{1}{4K^2} [ C\ti{12}^{(00)}, {{\cal J}_L}\ti{2}] \delta_{xy} - \frac{1}{2K}
C\ti{12}^{(00)} \delta'_{xy}\\
&- \frac{1}{4 K} m \Bigl(\mu^{-1} [C\ti{12}^{(00)},
(g^{-1} \Omega g)\ti{2}]
+ \lambda^{-1} [C\ti{12}^{(11)},
(g^{-1} \Omega g)\ti{2}]
\Bigr)
\delta_{xy},
\end{aligned}\end{equation}
from which, using \eqref{defofr12}, we immediately see that the terms proportional to $\delta'_{xy}$ in \eqref{cont limit lattice rs} and \eqref{pbLLcheck} agree.
To check the terms proportional to $\delta_{xy}$, we compute
\begin{equation}
[r\ti{12}(\lambda,\mu) + s\ti{12}, L\ti{1}] + [r\ti{12}(\lambda,\mu) - s\ti{12}, L\ti{2}],
\end{equation}
and confirm that it reproduces the relevant terms on the r.h.s. \eqref{pbLLcheck}.
Projecting the property~\eqref{property Casimir} onto $\alg{f}^{(i)} \otimes \alg{f}^{(j)}$ for $i,j=0,1$, we find the useful relations %v2
\begin{equation}\begin{gathered}\label{eq:casimirprop}
[C\ti{12}^{(00)},\alga\ti{1}+\alga\ti{2}] = [C\ti{12}^{(11)},\alga\ti{1}+\alga\ti{2}] = 0 ~, \qquad \alga \in \alg{f}^{(0)} ~,
\\
[C\ti{12}^{(00)},\alga\ti{1}] + [C\ti{12}^{(11)},\alga\ti{2}] = [C\ti{12}^{(11)},\alga\ti{1}] + [C\ti{12}^{(00)},\alga\ti{2}] = 0 ~, \qquad \alga \in \alg{f}^{(1)} ~.
\end{gathered}\end{equation}

Since ${\cal J}_L \in \f^{(0)}$, we have that
\begin{flalign*}
\ \ & \Bigl( [r\ti{12}(\lambda,\mu) + s\ti{12}, {{\cal J}_L}\ti{1}(x)]
+ [r\ti{12}(\lambda,\mu) - s\ti{12}, {{\cal J}_L}\ti{2}(y)]\Bigr)
\delta_{xy} & \\
& = [s\ti{12}, \frac{1}{2K}({{\cal J}_L}\ti{1}(x)-{{\cal J}_L}\ti{2}(y))]\delta_{xy}
= - \frac{1}{K}[s\ti{12}, {{\cal J}_L}\ti{2}(x)]\delta_{xy}
=- \frac{1}{4 K^2}[C\ti{12}^{(00)}, {{\cal J}_L}\ti{2}(x)]\delta_{xy}. &
\end{flalign*}
On the other hand, using that $\Omega \in \alg{f}^{(1)}$ and $g^{-1}\Omega g \in \alg{f}^{(1)}$, we have that
\begin{flalign*}
\ \ &\Bigl( [r\ti{12}(\lambda,\mu) + s\ti{12},- \ha \lambda
m \Omega\ti{1}] + [r\ti{12}(\lambda,\mu) - s\ti{12},- \ha \mu
m \Omega\ti{2}]\Bigr)\delta_{xy} &
\\
&= \frac{-m}{4 K (\mu^2 -\lambda^2)}
\Bigl(
[\mu^2 C\ti{12}^{(00)} + \lambda \mu C\ti{12}^{(11)}, \lambda
\Omega\ti{1}]
+[\lambda^2 C\ti{12}^{(00)} + \lambda \mu C\ti{12}^{(11)},
\mu
\Omega\ti{2}]
\Bigr)\delta_{xy} & \\
&=\frac{-m\mu\lambda}{4 K (\mu^2 -\lambda^2)}
\Bigl(
[-\mu C\ti{12}^{(11)} - \lambda C\ti{12}^{(00)},
\Omega\ti{2}]
+[\lambda C\ti{12}^{(00)} + \mu C\ti{12}^{(11)},
\Omega\ti{2}]
\Bigr)\delta_{xy}
=0, &
\end{flalign*}
and
\begin{flalign*}
\ \ &\Bigl( [r\ti{12}(\lambda,\mu) + s\ti{12},\ha\lambda^{-1}
m (g^{-1}\Omega g)\ti{1}] + [r\ti{12}(\lambda,\mu) - s\ti{12}, \ha \mu^{-1}
m (g^{-1}\Omega g)\ti{2}]\Bigr)\delta_{xy}
& \\
& =\frac{m}{4 K (\mu^2 -\lambda^2)}
\Bigl(
[\mu^2 C\ti{12}^{(00)} + \lambda \mu C\ti{12}^{(11)}, \lambda^{-1}
(g^{-1}\Omega g)\ti{1}]
+[\lambda^2 C\ti{12}^{(00)} + \lambda \mu C\ti{12}^{(11)}, \mu^{-1}
(g^{-1}\Omega g)\ti{2}]
\Bigr)\delta_{xy} & \\
& =\frac{m}{4 K (\mu^2 -\lambda^2)}
\Bigl(
[-\frac{\mu^2}{\lambda} C\ti{12}^{(11)} - \mu C\ti{12}^{(00)},
(g^{-1}\Omega g)\ti{2}]
+[\frac{\lambda^2}{\mu} C\ti{12}^{(00)} + \lambda C\ti{12}^{(11)},
(g^{-1}\Omega g)\ti{2}]
\Bigr)\delta_{xy} & \\
&=-\frac{m}{4 K} [\mu^{-1}C\ti{12}^{(00)} +
\lambda^{-1}C\ti{12}^{(11)} , (g^{-1}\Omega g)\ti{2}]\delta_{xy}. &
\end{flalign*}
Substituting~\eqref{defofr12} and~\eqref{eq:laxham} into~\eqref{cont limit lattice rs} and using the results above, we find~\eqref{pbLLcheck} as claimed.

\medskip

As explained in \cite{Delduc:2012qb}, the matrix $r+s$ is a solution of the Classical
Yang-Baxter Equation (CYBE).
More precisely, it is a twisted ${\cal R}$-matrix of the loop algebra of $\f$, where the twist
function corresponds to the generalised Faddeev-Reshetikhin procedure.
In appendix \ref{sss context alleviating} we demonstrate that $r$ satisfies the equation
\begin{equation}\begin{split} \label{CYBE type ra text}
[r\ti{12}(\lambda_1,\lambda_2),
r\ti{13}(\lambda_1, \lambda_3)]
+
[r\ti{12}(\lambda_1, \lambda_2),
r\ti{23}(\lambda_2, \lambda_3)]
+
[r\ti{13}(\lambda_1, \lambda_3),
r\ti{23}(\lambda_2, \lambda_3)] \qquad
\\
= \frac{1}{16 K^2} [C^{(00)}\ti{12},C^{(00)}\ti{13}].
\end{split}\end{equation}

\section{Lattice regularisation}\label{sec4}

For a non-ultralocal Poisson bracket of $r/s$ form, the Poisson bracket between two transition
matrices
\beqz
\{ \overleftarrow{U}(x_1,x_2,-L(\lambda))\ti{1},
\overleftarrow{U}(y_1,y_2,-L(\mu))\ti{2}\}
\eeqz
is ill-defined when either of the two points $(x_1,x_2)$
coincides with either of the two points $(y_1,y_2)$
\cite{Maillet:1985fn,Maillet:1985ek}.
One way to proceed is to use the prescription put forward in \cite{Maillet:1985fn,Maillet:1985ek}, in which the ill-defined
Poisson bracket $\{ {T_\ell}\ti{1}(\lambda), {T_\ell}\ti{2}(\mu) \} $ is set to
\beq
-[r\ti{12}(\lambda,\mu) , {T_\ell}\ti{1}(\lambda),{T_\ell}\ti{2}(\mu)].
\eeq
However, since $r$ itself is neither a solution of the CYBE nor a solution of the modified Classical Yang-Baxter Equation (mCYBE), the Jacobi identity is not satisfied.

As shown in \cite{Delduc:2012qb}, for the symmetric space Sine-Gordon models
the non-ultralocality is mild.
This means that there exists a more suitable prescription for defining the values of Poisson brackets of transition matrices that are ill-defined, which we review below.
We start by recalling some results of \cite{Delduc:2012qb} (see also \cite{SemenovTianShansky:1994dm,SemenovTianShansky:1995ha}).

\paragraph{Quadratic lattice algebra.}
The Poisson bracket of $L(x,\lambda)$ can be understood as stemming from an $abcd$ lattice quadratic algebra.
Let $\alpha \in \text{End}\, \g^{\mathbb{C}}$ be any skew-symmetric
solution of the mCYBE on $\g$
\beq
\forall \alga, \algb \in \g^{\mathbb{C}}, \qquad
[\alpha \alga,\alpha \algb] - \alpha \bigl( [\alpha \alga,\algb] +[\alga, \alpha \algb] \bigr) = - \xi^2 [\alga,\algb], \label{mybm}
\eeq
where for now we keep $\xi$ as a free parameter.
We extend $\alpha$ from $\g^{\mathbb{C}}$ to $\f^{\mathbb{C}}$ by letting it act trivially
on $\f^{(1)\mathbb{C}}$, i.e. $\alpha : \f^{(1)\mathbb{C}} \to 0$.
Since the operator $\alpha$ is a solution of the mCYBE on the finite Lie algebra $\g$,
the associated kernel $\alpha\ti{12}$ does not depend on the spectral parameter.

We define
\beq \label{def abcd}
a\ti{12} = -(r + \alpha)\ti{12}, \qquad b\ti{12} = (s + \alpha)\ti{12},
\qquad c\ti{12} = (s - \alpha)\ti{12}, \qquad d\ti{12}= -(r - \alpha)\ti{12}.
\eeq
The matrices $a$, $b$, $c$ and $d$ are not independent since
\beq \label{spec rela abcd}
a +b = d+c.
\eeq
Moreover, they satisfy the relations
\begin{subequations} \label{cybe abcd}
\begin{align}
[a\ti{12}, a\ti{13}] + [a\ti{12}, a\ti{23}] +[a\ti{13}, a\ti{23}] &=(\frac{1}{16K^2}-\xi^2) [C\ti{12}^{(00)},C\ti{13}^{(00)}]. \label{cybe abcd a} \\
[d\ti{12}, d\ti{13}] + [d\ti{12}, d\ti{23}] +[d\ti{13}, d\ti{23}] &=(\frac{1}{16K^2}-\xi^2) [C\ti{12}^{(00)},C\ti{13}^{(00)}], \label{cybe abcd d} \\
[a\ti{12},c\ti{13}] +[a\ti{12},c\ti{23}]+[c\ti{13},c\ti{23}]
&=(\frac{1}{16K^2}-\xi^2) [C\ti{12}^{(00)},C\ti{13}^{(00)}],\\
[d\ti{12},b\ti{13}] +[d\ti{12},b\ti{23}]+[b\ti{13},b\ti{23}]
&=(\frac{1}{16K^2}-\xi^2) [C\ti{12}^{(00)},C\ti{13}^{(00)}].
\end{align}
\end{subequations}
In these equations, and others below, we suppress the dependence on the spectral parameters whenever there is no ambiguity.
The proof of these equations is given in appendix \ref{sss context alleviating}.

We now consider a one-dimensional lattice of points $x_n$ with lattice spacing $\Delta = x_{n+1} - x_n$.
At each lattice site we introduce a quantity
${\cal L}^n(\lambda,t)$
and define the quadratic lattice algebra \cite{Freidel:1991jx,Freidel:1991jv}
\begin{multline} \label{lattice algebra 2}
\{ {\cal L}^n\ti{1}(\lambda), {\cal L}^m\ti{2}(\mu) \} =
a\ti{12}(\lambda,\mu) {\cal L}^n\ti{1}(\lambda) {\cal L}^m\ti{2}(\mu) \delta_{m n}
- {\cal L}^n\ti{1}(\lambda) {\cal L}^m\ti{2}(\mu) d\ti{12}(\lambda,\mu)\delta_{m n} \\
+ {\cal L}^n\ti{1}(\lambda) b\ti{12}{\cal L}^m\ti{2}(\mu) \delta_{m+1, n}
- {\cal L}^m\ti{2}(\mu) c\ti{12}{\cal L}^n\ti{1}(\lambda) \delta_{m, n+1},
\end{multline}
where in the case at hand the matrices $b\ti{12}$ and $c\ti{12}$ are independent of the spectral parameter.
On the lattice, the notion of non-ultralocality is more transparent than in the continuum.
The terms proportional to $a\ti{12}$ and $d\ti{12}$ in the Poisson bracket are ultralocal since they vanish except when ${\cal L}^n\ti{1}(\lambda)$ and ${\cal L}^m\ti{2}(\mu)$ are at the same site, i.e. $n=m$, while the terms proportional to $b\ti{12}$ and $c\ti{12}$ are non-ultralocal
since they vanish except when ${\cal L}^n\ti{1}(\lambda)$ and ${\cal L}^m\ti{2}(\mu)$ are at adjacent sites, i.e. $n=m\pm 1$.
If we take $\xi^2=1/(16 K^2)$, then the properties \eqref{cybe abcd} ensure \cite{Freidel:1991jx,Freidel:1991jv} that the bracket \eqref{lattice algebra 2} satisfies the Jacobi identity.

\paragraph{Continuum limit.}
The standard continuum limit is given by setting
\beq
{\cal L}^n(\lambda,t) =
\id - \Delta \, L(\lambda,x_n,t) + O(\Delta^2),
\eeq
and taking the lattice spacing $\Delta \to 0$.
Let $x=x_n$ and $x'=x_m$. In the continuum limit
\begin{align*}
\delta_{mn} & \stackrel[\Delta \to 0]{}{\sim} \Delta \delta(x-x'), \qquad
\delta_{m+1, n} - \delta_{mn}
\stackrel[\Delta \to 0]{}{\sim}
- \Delta^2
\partial_x \delta(x-x').
\end{align*}
Taking the continuum limit on the l.h.s. of \eqref{lattice algebra 2} we find
\beq
\Delta^2 \{ L\ti{1}(\lambda,x), L\ti{2}(\mu,x') \} +\Order(\Delta^3).
\eeq
We now list the terms obtained on the r.h.s. up to $\Order(\Delta^2)$.
The first set of terms comes from the leading contribution in the expansion
of ${\cal L}^n\ti{1}$ and ${\cal L}^m\ti{2}$, i.e. setting both equal to $\id$ in~\eqref{lattice algebra 2}.
Their sum is
\begin{equation}
\begin{aligned}
& (a\ti{12}-d\ti{12}+b\ti{12}-c\ti{12}) \delta_{mn}+ b\ti{12}(\delta_{m+1,n} -\delta_{mn})
+c\ti{12}(\delta_{mn}-\delta_{m,n+1})\\
& \quad = b\ti{12}(\delta_{m+1,n} -\delta_{mn})
+c\ti{12}(\delta_{mn}-\delta_{m,n+1}) \\
& \quad \sim- \Delta^2 (b\ti{12}+c\ti{12}) \partial_x \delta(x-x').
\end{aligned}\end{equation}
Note that the (standard) continuum limit would not be defined without the relation \eqref{spec rela abcd}.

The second set of terms are those linear in $L$ originating from the ultralocal part of~\eqref{lattice algebra 2}
\begin{equation}\begin{aligned}
& - \Delta \delta_{mn} ( a\ti{12}(L\ti{1}(\lambda,x_n) + {L}\ti{2}(\mu,x_m))
- ({L}\ti{1}(\lambda,x_n) + {L}\ti{2}(\mu,x_m)) d\ti{12} )\\
&\quad \sim - \Delta^2 \delta(x-x') \bigl( a\ti{12} (L\ti{1}(\lambda,x)
+ L\ti{2}(\mu,x) ) - (L\ti{1}(\lambda,x)
+ L\ti{2}(\mu,x) ) d\ti{12} \bigr).
\end{aligned}\end{equation}
The third set is formed by the terms linear in $L$ coming from the non-ultralocal part of~\eqref{lattice algebra 2}.
To analyse these, let us focus on the term
\begin{equation}\begin{aligned}
-\Delta {L}\ti{1}(\lambda,x_n) b\ti{12} \delta_{m+1,n}
&= - \Delta {L}\ti{1}(\lambda,x_n) b\ti{12} \delta_{mn}
- \Delta {L}\ti{1}(\lambda,x_n) (\delta_{m+1,n} -\delta_{mn}) \\
& \sim
-\Delta^2 L\ti{1}(\lambda,x) b\ti{12} \delta(x-x') + \Order(\Delta^3).
\end{aligned}\end{equation}
Treating the other three terms similarly, the third set of terms contributing at $\Order(\Delta^2)$ are
\beq
- \Delta^2 \delta(x-x')
\bigl(
L\ti{1}(\lambda,x) b\ti{12} + b\ti{12} L\ti{2}(\mu,x)
- L\ti{2}(\mu,x) c\ti{12} - c\ti{12} L\ti{1}(\lambda,x)
\bigr).
\eeq

There are no further contributions at $\Order(\Delta^2)$.
Putting the three sets of terms together, we find that the continuum limit of the Poisson bracket \eqref{lattice algebra 2} is
\begin{align}
\{ L\ti{1}(\lambda,x), L\ti{2}(\mu,x') \}
&= -\Bigl( (a\ti{12}-c\ti{12}) L\ti{1}(\lambda,x)
+ (a\ti{12}+b\ti{12}) L\ti{2}(\mu,x') \nonumber \\
&\qquad - L\ti{1}(\lambda,x)(d\ti{12}-b\ti{12})
- L\ti{2}(\mu,x')(d\ti{12}+c\ti{12}) \Bigr) \delta(x-x') \nonumber \\
& \quad\, -(b\ti{12}+c\ti{12}) \partial_x \delta(x-x')\nonumber\\
&= - \Bigl( [ a\ti{12}-c\ti{12} , L\ti{1}(\lambda,x) ]
+ [a\ti{12}+b\ti{12}, L\ti{2}(\mu,x') ] \Bigr) \delta(x-x') \nonumber \\
&\quad\,-(b\ti{12}+c\ti{12}) \partial_x \delta(x-x'), \nonumber
\end{align}
where we have again used the relation \eqref{spec rela abcd}.
In our case, with the matrices $a$, $b$, $c$ and $d$ defined in terms of $r$, $s$ and $\alpha$ \eqref{def abcd}, this becomes
\begin{align}\label{cont limit lattice rs2}
\{ L\ti{1}(\lambda,x), L\ti{2}(\mu,x') \}
&= [r\ti{12}(\lambda,\mu), L\ti{1}(\lambda,x) + L\ti{2}(\mu,x')] \delta(x-x')
\\
& \quad + [s\ti{12} , L\ti{1}(\lambda,x) - L\ti{2}(\mu,x')] \delta(x-x') \nonumber
- 2 s\ti{12} \partial_x \delta(x-x').
\end{align}
Therefore, in the continuum limit, $\alpha$ does not contribute.
Equivalently, this implies that
\beq
\Delta^{-2}
\bigl(
- \alpha\ti{12} {\cal L}^n\ti{1} {\cal L}^m\ti{2} \delta_{m n}
- {\cal L}^n\ti{1} {\cal L}^m\ti{2} \alpha\ti{12}\delta_{m n}
+ {\cal L}^n\ti{1} \alpha\ti{12} {\cal L}^m\ti{2} \delta_{m+1, n}
+ {\cal L}^m\ti{2} \alpha\ti{12} {\cal L}^n\ti{1} \delta_{m, n+1}
\bigr)
\eeq
vanishes in the (standard) continuum limit.
It is in this sense that the $r/s$ Poisson bracket \eqref{cont limit lattice rs} of $L(\lambda,x)$
stems from the quadratic lattice algebra \eqref{lattice algebra 2} satisfied by
${\cal L}^n(\lambda)$.

\paragraph{Poisson bracket of transition matrices on the lattice.}

The transition matrix $T^{n,q}(\lambda) $ from $x_q$ to $x_{n+1}$ with $n \geq q$ is defined as the product of lattice Lax matrices on successive sites
\begin{equation}
T^{n,q} = {\cal L}^n {\cal L}^{n-1}\dots {\cal L}^{q+1}{\cal L}^{q},
\end{equation}
where we have suppressed the dependence on the spectral parameter.
In appendix~\ref{ttpbproof} we show that the Poisson bracket $\{T\ti{1}^{n,q}(\lambda) ,
T\ti{2}^{m,p}(\mu) \}$ is given by
\begin{equation}\begin{split}\label{eq:ttpb}
\{T\ti{1}^{n,q},T\ti{2}^{m,p} \} = &
\sum_{r=1}^{n-q}T\ti{1}^{n,q+r}\big((a\ti{12} + b\ti{12})T\ti{2}^{m,p}\delta_{m+1,q+r}
-
T\ti{2}^{m,p}(d\ti{12} + c\ti{12})\delta_{p,q+r}\big)T\ti{1}^{q+r-1,q}
\\ + &
\sum_{s=1}^{m-p} T\ti{2}^{m,p+s}\big((a\ti{12} - c\ti{12})T\ti{1}^{n,q}\delta_{p+s,n+1}
-
T\ti{1}^{n,q}(d\ti{12} - b\ti{12})\delta_{p+s,q}\big)T\ti{2}^{p+s-1}
\\ + &
a\ti{12} T\ti{1}^{n,q} T\ti{2}^{m,p} \delta_{m,n} - T\ti{1}^{n,q} T\ti{2}^{m,p} d\ti{12} \delta_{p,q}
+ T\ti{1}^{n,q} b\ti{12} T\ti{2}^{m,p} \delta_{m+1,q} - T\ti{2}^{m,p} c\ti{12} T\ti{1}^{n,q} \delta_{p,n+1} ,
\end{split}\end{equation}
where we have again suppressed the dependence on the spectral parameters.
If we set $q=n$ and $p=m$, we have $T\ti{1}^{n,n} = {\cal L}\ti{1}^n$ and $T\ti{2}^{m,m} = {\cal L}\ti{2}^m$.
There are then no terms in the sums in~\eqref{eq:ttpb} and we are just left with the final line, recovering~\eqref{lattice algebra 2} as expected.

\begin{figure}[t]
\begin{center}
\small
\begin{tikzpicture}[scale=0.8,
*/.tip={Circle[sep=-1.196825pt -1.595769]},
o/.tip={Circle[sep=-1.196825pt -1.595769, open]},
]
%\node at (-8,0) {Lattice:};
%\node at (-3,0) {$\times$};
%\node at (-2,0) {$\times$};
%\node at (-1,0) {$\times$};
%\node at (0,0) {$\times$};
%\node at (1,0) {$\times$};
%\node at (2,0) {$\times$};
%\node at (3,0) {$\times$};
%\node at (4,0) {$\times$};
\node at (-8,-1) {Case 1};
\node at (-8,-1.8) {a) $p=n+1$:};
\node at (-8,-2.6) {b) $m=q-1$:};
\node at (-3,-1) {$\times$};
\node at (-2,-1) {$\times$};
\node at (-1,-1) {$\times$};
\node at (0,-1) {$\times$};
\node at (1,-1) {$\times$};
\node at (2,-1) {$\times$};
\node at (3,-1) {$\times$};
\node at (4,-1) {$\times$};
\draw[<-*,very thick] (-1.5,-1.8)--(0,-1.8);
\draw[*->,very thick] (1,-1.8)--(2.5,-1.8);
\draw[<-*,very thick] (-1.5,-2.6)--(0,-2.6);
\draw[*->,very thick] (1,-2.6)--(2.5,-2.6);
\node at (-2,-1.8) {$T\ti{1}^{n,q}$};
\node at (3.4,-1.8) {$T\ti{2}^{m,n+1}$};
\node at (-2.2,-2.6) {$T\ti{2}^{q-1,p}$};
\node at (3.1,-2.6) {$T\ti{1}^{n,q}$};
\node at (-8,-5) {Case 2:};
\node at (-8,-5.8) {($p=q$)};
\node at (-3,-5) {$\times$};
\node at (-2,-5) {$\times$};
\node at (-1,-5) {$\times$};
\node at (0,-5) {$\times$};
\node at (1,-5) {$\times$};
\node at (2,-5) {$\times$};
\node at (3,-5) {$\times$};
\node at (4,-5) {$\times$};
\draw[*->,very thick] (-1,-4.7)--(2.7,-4.7);
\draw[*->,very thick] (-1,-5.3)--(2.5,-5.3);
\node at (0,-4.2) {$T\ti{1}^{n,q}$};
\node at (0,-5.8) {$T\ti{2}^{m,q}$};
\node at (-8,-8) {Case 3:};
\node at (-8,-8.8) {($m=n$)};
\node at (-3,-8) {$\times$};
\node at (-2,-8) {$\times$};
\node at (-1,-8) {$\times$};
\node at (0,-8) {$\times$};
\node at (1,-8) {$\times$};
\node at (2,-8) {$\times$};
\node at (3,-8) {$\times$};
\node at (4,-8) {$\times$};
\draw[<-*,very thick] (-1.7,-7.7)--(2,-7.7);
\draw[<-*,very thick] (-1.5,-8.3)--(2,-8.3);
\node at (1,-7.2) {$T\ti{1}^{n,q}$};
\node at (1,-8.8) {$T\ti{2}^{n,p}$};
\node at (-8,-11) {Case 4:};
\node at (-8,-11.8) {($p=q$ and $m=n$)};
\node at (-3,-11) {$\times$};
\node at (-2,-11) {$\times$};
\node at (-1,-11) {$\times$};
\node at (0,-11) {$\times$};
\node at (1,-11) {$\times$};
\node at (2,-11) {$\times$};
\node at (3,-11) {$\times$};
\node at (4,-11) {$\times$};
\draw[*-*,very thick] (-1,-10.7)--(2,-10.7);
\draw[*-*,very thick] (-1,-11.3)--(2,-11.3);
\node at (0.5,-10.2) {$T\ti{1}^{n,q}$};
\node at (0.5,-11.8) {$T\ti{2}^{n,q}$};
\node at (0.55,-11) {$\dots$};
\end{tikzpicture}
\end{center}
\caption{Those cases for which the Poisson bracket of two transfer matrices $\{T\ti{1}^{n,q} ,T\ti{2}^{m,p} \}$ depends on $\alpha$.
Recall that for $T^{n,q}$ we have $n \geq q$.
The lattice sites are represented by $\times$, the dots denote that the transfer matrix ends at this site, while the arrows denote that the transfer matrix can carry on arbitrarily in the corresponding direction.}\label{Three cases}
\end{figure}
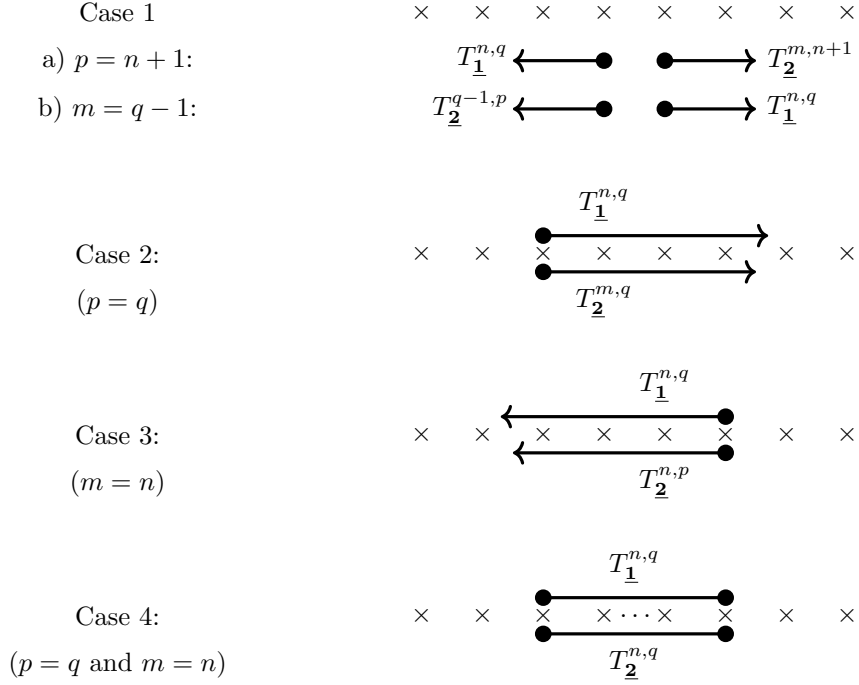

Recalling the definitions~\eqref{def abcd}, we note that the first two lines of \eqref{eq:ttpb} are independent of $\alpha$. Furthermore, the four terms on the final line have different structures. They cannot combine in order to cancel the dependence on $\alpha$.
Thus, the Poisson bracket $\{T\ti{1}^{n,q},T\ti{2}^{m,p} \}$ is independent of $\alpha$ except when
at least one of these four terms does not vanish, for instance when $m=n$.
In figure \ref{Three cases} we have indicated those cases for which the Poisson bracket does depend on $\alpha$.

Our interest is in the case $n=m$ and $q=p$.
In this case, it is straightforward to see that the sums in~\eqref{eq:ttpb} do not contribute since the Kronecker deltas always vanish.
Moreover, since $n\geq q$ only the terms proportional to $a\ti{12}$ and $d\ti{12}$ on the final line survive, and we are left with
\beq \label{pb TT lattice}
\{ T^{n,q}\ti{1}(\lambda), T^{n,q}\ti{2}(\mu) \}
= a\ti{12}(\lambda,\mu) T^{n,q}\ti{1}(\lambda), T^{n,q}\ti{2}(\mu)
- T^{n,q}\ti{1}(\lambda), T^{n,q}\ti{2}(\mu) d\ti{12}(\lambda,\mu).
\eeq

\paragraph{Poisson bracket of transition matrices in the continuum.}
Since the $r/s$ structure stems from a lattice $abcd$
quadratic algebra, we may regularise the ill-defined Poisson bracket of
${T_\ell}(\lambda)$ with ${T_\ell}(\mu)$ by taking the continuum limit of \eqref{pb TT lattice}, which is simply
\begin{equation}\begin{aligned} \label{pb T T continuum}
\{ {T_\ell}\ti{1}(\lambda), {T_\ell}\ti{2}(\mu) \}
&= a\ti{12}(\lambda,\mu) {T_\ell}\ti{1}(\lambda) {T_\ell}\ti{2}(\mu)
- {T_\ell}\ti{1}(\lambda) {T_\ell}\ti{2}(\mu) d\ti{12}(\lambda,\mu),\\
&=-[r\ti{12}(\lambda,\mu) ,{T_\ell}\ti{1}(\lambda) {T_\ell}\ti{2}(\mu) ]
-\alpha\ti{12} {T_\ell}\ti{1}(\lambda) {T_\ell}\ti{2}(\mu)
- {T_\ell}\ti{1}(\lambda) {T_\ell}\ti{2}(\mu) \alpha\ti{12}.
\end{aligned}\end{equation}
We refer to this as the `lattice' regularisation. When $\xi^2=1/(16 K^2)$, the Jacobi identity for the bracket \eqref{pb T T continuum} is ensured since $a$ and $d$ satisfy the CYBE (see equations \eqref{cybe abcd a} and \eqref{cybe abcd d}).
The case $\xi=0$, i.e. $\alpha=0$, corresponds to the `standard' regularisation, for which the Jacobi identity is not satisfied.

\section{Poisson bracket of the subtracted monodromy}\label{sec5}

The Poisson bracket of the subtracted monodromy, defined in equation~\eqref{subT gsg 2}, follows from the Poisson bracket \eqref{pb T T continuum} and is given by
\beq \label{PB stell stell SSSG}
\{ {\st}\ti{1}(\lambda) , {\st}\ti{2}(\mu) \} =
\lim_{\ell \to + \infty}\big(
{a_\ell}\ti{12}(\lambda, \mu)
{\st_\ell}\ti{1}(\lambda) {\st_\ell}\ti{2}(\mu)
- {\st_\ell}\ti{1}(\lambda) {\st_\ell}\ti{2}(\mu) {d_\ell}\ti{12}(\lambda, \mu)\big),
\eeq
where
\begin{subequations}
\begin{align}
{a_\ell}\ti{12}(\lambda, \mu) &=
(\widehat{\Phi}^{0}\ti{1}(\lambda, \ell))^{-1}
(\widehat{\Phi}^{0}\ti{2}(\mu, \ell))^{-1}
a\ti{12}(\lambda, \mu)
\widehat{\Phi}^{0}\ti{1}(\lambda, \ell)
\widehat{\Phi}^{0}\ti{2}(\mu, \ell),\\
{d_\ell}\ti{12}(\lambda, \mu) &=
(\widehat{\Phi}^{0}\ti{1}(\lambda,-\ell))^{-1}
(\widehat{\Phi}^{0}\ti{2}(\mu,-\ell))^{-1}
d\ti{12}(\lambda, \mu)
\widehat{\Phi}^{0}\ti{1}(\lambda,-\ell)
\widehat{\Phi}^{0}\ti{2}(\mu,-\ell),
\end{align}
\end{subequations}
and we recall that the matrices $a$ and $d$ are defined in equation~\eqref{def abcd}.

To determine the $\ell \to +\infty$ limit in the Poisson bracket~\eqref{PB stell stell SSSG}, we recall that
$\widehat{\Phi}^0(\lambda,x) = \exp\bigl( k(\lambda) x \Omega \bigr)$.
Therefore, making the assumption on the algebraic structure outlined at the end of subsection~\ref{sec:alg}, we decompose $r$ and our choice for $\alpha$ in terms of the eigenspaces $\omega^\pm$, $\alg{h}$ and $\alg{a}$ of the adjoint action of $\Omega$.
To do so, we introduce a suitable basis
\unskip\footnote{For the spheres, with $F = SO(N+1)$, the index $a$ runs from 1 to $N-1$.}
$\{\omega^\pm_a\}$ of $\omega^\pm$, in which
\begin{subequations}
\begin{align}
C\ti{12} &= C\ti{12}^\h + C\ti{12}^{\alg{a}}
+{\omega^+_a}\ti{1} {\omega^-_a}\ti{2}
+{\omega^-_a}\ti{1} {\omega^+_a}\ti{2},
\\\label{c00w}
C\ti{12}^{(00)} &= C^\h\ti{12} +\frac{1}{2}
(
{\omega^+_a}\ti{1}+{\omega^-_a}\ti{1}
)(
{\omega^+_a}\ti{2}+{\omega^-_a}\ti{2}
),\\\label{c11w}
C\ti{12}^{(11)} &= C\ti{12}^{\alg{a}}
- \frac{1}{2}
(
{\omega^+_a}\ti{1}-{\omega^-_a}\ti{1}
)
(
{\omega^+_a}\ti{2}-{\omega^-_a}\ti{2}
).
\end{align}
\end{subequations}
It will also be useful to introduce the kernel $\rho\ti{12}$ of $\ad_\Omega$
\begin{equation}
\rho\ti{12} = [\Omega\ti{1},C\ti{12}] = i ( {\omega^+_a}\ti{1} {\omega^-_a}\ti{2} -{\omega^-_a}\ti{1} {\omega^+_a}\ti{2}),
\end{equation}
where we note that $\rho\ti{12}$ is skew-symmetric, $\rho\ti{12}=-\rho\ti{21}$.

Now using that $\widehat{\Phi}^0(\lambda,x)=\exp(k(\lambda) x \Omega)$, we have the following relations
\begin{subequations} \label{rot list}
\begin{align}
(\widehat{\Phi}^0\ti{1}(\lambda,x))^{-1}
(\widehat{\Phi}^0\ti{2}(\mu,x))^{-1} \,
C^\h\ti{12} \,
\widehat{\Phi}^0\ti{1}(\lambda,x)
\widehat{\Phi}^0\ti{2}(\mu,x)
&= C^\h\ti{12} \label{rotCh},\\
(\widehat{\Phi}^0\ti{1}(\lambda,x))^{-1}
(\widehat{\Phi}^0\ti{2}(\mu,x))^{-1} \,
C\ti{12}^{\alg{a}} \,
\widehat{\Phi}^0\ti{1}(\lambda,x)
\widehat{\Phi}^0\ti{2}(\mu,x)
&= C\ti{12}^{\alg{a}},\label{rotomega}\\
(\widehat{\Phi}^0\ti{1}(\lambda,x))^{-1}
(\widehat{\Phi}^0\ti{2}(\mu,x))^{-1} \,
{\omega^+_a}\ti{1}{\omega^-_b}\ti{2} \,
\widehat{\Phi}^0\ti{1}(\lambda,x)
\widehat{\Phi}^0\ti{2}(\mu,x)
&= e^{-i (k(\lambda)-k(\mu))x }
{\omega^+_a}\ti{1}{\omega^-_b}\ti{2},\label{55c}\\
(\widehat{\Phi}^0\ti{1}(\lambda,x))^{-1}
(\widehat{\Phi}^0\ti{2}(\mu,x))^{-1} \,
{\omega^-_a}\ti{1}{\omega^+_b}\ti{2} \,
\widehat{\Phi}^0\ti{1}(\lambda,x)
\widehat{\Phi}^0\ti{2}(\mu,x)
&= e^{i (k(\lambda)-k(\mu))x }
{\omega^-_a}\ti{1}{\omega^+_b}\ti{2},\label{55d}\\
(\widehat{\Phi}^0\ti{1}(\lambda,x))^{-1}
(\widehat{\Phi}^0\ti{2}(\mu,x))^{-1} \,
{\omega^+_a}\ti{1}{\omega^+_b}\ti{2} \,
\widehat{\Phi}^0\ti{1}(\lambda,x)
\widehat{\Phi}^0\ti{2}(\mu,x)
&= e^{-i (k(\lambda)+k(\mu))x }
{\omega^+_a}\ti{1}{\omega^+_b}\ti{2},\\
(\widehat{\Phi}^0\ti{1}(\lambda,x))^{-1}
(\widehat{\Phi}^0\ti{2}(\mu,x))^{-1} \,
{\omega^-_a}\ti{1}{\omega^-_b}\ti{2} \,
\widehat{\Phi}^0\ti{1}(\lambda,x)
\widehat{\Phi}^0\ti{2}(\mu,x)
&= e^{i (k(\lambda)+k(\mu))x }
{\omega^-_a}\ti{1}{\omega^-_b}\ti{2}.
\end{align}
\end{subequations}

\subsection{Contribution of \texorpdfstring{$r$}{r}}

Using the expressions~\eqref{c00w} and \eqref{c11w}, the matrix $r$~\eqref{defofr12} can be written as
\begin{multline}
r\ti{12}(\lambda,\mu)
=
\frac{1}{4K}
\Bigl(
\frac{\mu^2 +\lambda^2}{\mu^2 -\lambda^2}
C^\h\ti{12}
+ \frac{2 \lambda \mu}{\mu^2 -\lambda^2}
C\ti{12}^{\alg{a}} \\
- \frac{1}{2} \frac{\lambda-\mu}{\lambda+\mu}
({\omega^+_a}\ti{1}{\omega^+_a}\ti{2}
+{\omega^-_a}\ti{1}{\omega^-_a}\ti{2})
-\frac{1}{2} \frac{\lambda+\mu}{\lambda-\mu}
({\omega^+_a}\ti{1}{\omega^-_a}\ti{2}
+{\omega^-_a}\ti{1}{\omega^+_a}\ti{2})
\Bigr).
\end{multline}
In order to compute the limit in equation~\eqref{PB stell stell SSSG}, we restrict to $\lambda>0$ and $\mu>0$, with the extension to any $\lambda \in \mathbb{R}$ and $\mu \in \mathbb{R}$ explained in subsection \ref{sec:fullext}.
We will also assume that the analytical properties of $\st_\ell$ are such that the limit can be taken in the sense of the Cauchy principal value, or more precisely, that we may apply the results
\unskip\footnote{These results rely on the observation that $k(\lambda) = \frac{m}{2}(\lambda -\lambda^{-1})$ implies that $\Im k(\lambda) = \frac{m}{2}(1+ |\lambda|^{-2}) \Im \lambda$,
hence $\Im k(\lambda)$ has the same sign as $\Im \lambda$. In particular it has a fixed
sign in the lower and upper complex half-planes.}
\cite{fadtakbook87,Babebook}
\begin{subequations} \label{formula pp as limit}
\begin{align}
\lim\limits_{\ell \to +\infty} \PP \frac{1}{\lambda-\mu}
\exp\Bigl( \pm i \ell (k(\lambda) - k(\mu)) \Bigr) &= \pm i \pi \delta(\lambda-\mu),\\
\lim\limits_{\ell \to +\infty} \PP \frac{1}{\lambda + \mu}
\exp\Bigl( \pm i \ell (k(\lambda) + k(\mu)) \Bigr) &= \pm i \pi \delta(\lambda+\mu) =0,
\end{align}
\end{subequations}
where the final equality follows since we take both $\lambda$ and $\mu$ to be positive.

Therefore, we are interested in the limits
\begin{equation}\begin{gathered}\label{ADlim}
\widetilde{r}^\pm\ti{12}(\lambda,\mu) = -\lim_{\ell\to+\infty}(\widehat{\Phi}^{0}\ti{1}(\lambda, \pm \ell))^{-1}
(\widehat{\Phi}^{0}\ti{2}(\mu, \pm \ell))^{-1}
r\ti{12}(\lambda, \mu)
\widehat{\Phi}^{0}\ti{1}(\lambda, \pm \ell)
\widehat{\Phi}^{0}\ti{2}(\mu, \pm \ell).
\end{gathered}\end{equation}
Using the results~\eqref{rot list} and \eqref{formula pp as limit}, we find
\begin{equation}\begin{aligned}\label{ADlim2}
\widetilde{r}^\pm\ti{12}(\lambda,\mu) &=
-\frac{1}{4K} \Bigl(\varphi^+(\lambda,\mu) C\ti{12}^\h +
\varphi^-(\lambda,\mu) C\ti{12}^{\alg{a}}
\pm \lambda \pi \rho\ti{12} \delta(\lambda-\mu)
\Bigr),
\end{aligned}\end{equation}
where we have defined
\begin{equation}
\varphi^\pm(\lambda,\mu) =-\frac12\Bigl(
\PP \frac{\lambda+\mu}{\lambda-\mu} \pm \frac{\lambda-\mu}{\lambda+\mu}
\Bigr). \label{def phipm}
\end{equation}

Here, to make sense of the $\lambda - \mu \to 0$ limit, we have taken the Cauchy principal value also in the terms proportional to $C\ti{12}^\h$ and $C\ti{12}^{\alg{a}}$.
The justification follows an argument given in \cite{Sklyanin:1980ij}.
As we shall see below, the contribution stemming from $\alpha\ti{12}$ is not singular when $\lambda -\mu \to 0$.
Therefore, in the limit where $\lambda-\mu \to 0$, the
apparently divergent term in the Poisson bracket is
\begin{equation}\begin{aligned}
\lim_{\lambda - \mu \to 0} &\big(
\widetilde{r}^+\ti{12}(\lambda,\mu) \st(\lambda)\ti{1} \st(\mu) \ti{2}
-\st(\lambda)\ti{1} \st(\mu) \ti{2} \widetilde{r}^-\ti{12}(\lambda,\mu)\big)
\\
\sim
& -\frac{\lambda}{4K}
\Bigl( -
\PP \frac{1}{\lambda-\mu}
C\ti{12}^\h
- \PP \frac{1}{\lambda-\mu}
C\ti{12}^{\alg{a}}
+ \pi \rho\ti{12} \delta(\lambda-\mu)
\Bigr)\st(\lambda)\ti{1} \st(\lambda) \ti{2}
\\
& + \frac{\lambda}{4K} \st(\lambda)\ti{1} \st(\lambda) \ti{2}
\Bigl(
- \PP \frac{1}{\lambda-\mu}
C\ti{12}^\h
- \PP \frac{1}{\lambda-\mu}
C\ti{12}^{\alg{a}}
- \pi \rho\ti{12} \delta(\lambda-\mu)\Bigr)
\\
= &-\frac{\lambda}{4K}
\Bigl( -
\PP \frac{1}{\lambda-\mu}
C\ti{12} +
\PP \frac{1}{\lambda-\mu}
( {\omega^+_a}\ti{1} {\omega^-_a}\ti{2}
+{\omega^-_a}\ti{1} {\omega^+_a}\ti{2}) \\
& \hspace{114pt}+i \pi \delta(\lambda-\mu)( {\omega^+_a}\ti{1} {\omega^-_a}\ti{2}
-{\omega^-_a}\ti{1} {\omega^+_a}\ti{2})
\Bigr) \st(\lambda)\ti{1} \st(\lambda) \ti{2} \\
&+\frac{\lambda}{4K}
\st(\lambda)\ti{1} \st(\lambda) \ti{2}
\Bigl( -
\PP \frac{1}{\lambda-\mu}
C\ti{12} +
\PP \frac{1}{\lambda-\mu}
( {\omega^+_a}\ti{1} {\omega^-_a}\ti{2}
+{\omega^-_a}\ti{1} {\omega^+_a}\ti{2}) \\
& \hspace{167pt}-i \pi \delta(\lambda-\mu)( {\omega^+_a}\ti{1} {\omega^-_a}\ti{2}
-{\omega^-_a}\ti{1} {\omega^+_a}\ti{2})
\Bigr)
\\
= & \frac{\lambda}{4K} \PP \frac{1}{\lambda-\mu} [C\ti{12}, \st(\lambda)\ti{1} \st(\lambda) \ti{2} ]
\\ &
-\frac{\lambda}{4K}\lim\limits_{\epsilon \to 0^+}
\Bigl(
\frac{1}{\lambda-\mu - i \epsilon} {\omega^+_a}\ti{1} {\omega^-_a}\ti{2}
+
\frac{1}{\lambda-\mu +i \epsilon} {\omega^-_a}\ti{1} {\omega^+_a}\ti{2}\Bigr)
\st(\lambda)\ti{1} \st(\lambda) \ti{2}
\\ &
+\frac{\lambda}{4K}\st(\lambda)\ti{1} \st(\lambda) \ti{2}\lim\limits_{\epsilon \to 0^+}
\Bigl(
\frac{1}{\lambda-\mu + i \epsilon} {\omega^+_a}\ti{1} {\omega^-_a}\ti{2}
+
\frac{1}{\lambda-\mu -i \epsilon} {\omega^-_a}\ti{1} {\omega^+_a}\ti{2}\Bigr),
\end{aligned}\end{equation}
where we have used the property
\beq
\lim\limits_{\epsilon \to 0^+} \frac{1}{x\mp i \epsilon}
= \PP\frac{1}{x} \pm i \pi \delta(x).
\eeq
Since
\beq
[C\ti{12}, \st(\lambda)\ti{1} \st(\lambda) \ti{2} ]=0,
\eeq
we see that taking the Cauchy principal value in the terms proportional
either to $C\ti{12}^\h$ or to $C\ti{12}^{\alg{a}}$ corresponds to a prescription
for the non-singular Poisson bracket $\{\st(\lambda)\ti{1},\st(\lambda)\ti{2}\}$.

\subsection{Choice and contribution of \texorpdfstring{$\alpha$}{alpha}}\label{sec:ccal}

We choose the matrix $\alpha$ to have the form
\beq
\alpha\ti{12} = \alpha^\h\ti{12} + \psi\ti{12}
\eeq
where $\alpha^\h\ti{12} \in \alg{h} \otimes \alg{h}$ is a skew-symmetric split $r$-matrix on $\h$, and $P^{\alg{h}}\ti{1} \psi\ti{12} = P^{\alg{h}}\ti{2} \psi\ti{12} = 0$.
Up to complexifying the algebra, we have checked that it is always possible to make such a choice for the spheres in appendix~\ref{spheres}.

In the limit $\ell \to \infty$, the terms in \eqref{PB stell stell SSSG}
proportional to $\alpha^\h$ simply give
\beq \label{lim coming alphah}
- \alpha^\h\ti{12} \st\ti{1}(\lambda) \st\ti{2}(\mu) - \st\ti{1}(\lambda) \st\ti{2}(\mu) \alpha^\h\ti{12}.
\eeq
Indeed, the matrix $\psi$ can be written as a linear combination of tensor products
of ${\omega^{\pm}_a}$. By definition there is no dependence on the spectral
parameters $\lambda$ and $\mu$.
The relations \eqref{rot list} then indicate that we
should take the limits of terms whose frequency of oscillation grows as $\ell \to \infty$.
Defining these limits in an analogous way to \eqref{formula pp as limit}, we see that they
are of the type $(\lambda\pm\mu) \delta(\lambda\pm \mu)$, hence vanish.
Therefore, only the $\alpha^\h$
contribution survives in the limit $\ell \to \infty$.

\subsection{Full result and extension to any real spectral parameters}\label{sec:fullext}

We have shown that for $\lambda >0$
and $\mu>0$ the Poisson bracket of the subtracted monodromy matrix is
\beq
\label{PB stell stell eq1}
\{ {\st}\ti{1}(\lambda) , {\st}\ti{2}(\mu) \} =
\widetilde{a}\ti{12}(\lambda, \mu)
{\st}\ti{1}(\lambda) {\st}\ti{2}(\mu)
- {\st}\ti{1}(\lambda) {\st}\ti{2}(\mu) \widetilde{d}\ti{12}(\lambda, \mu)
\eeq
with
\begin{equation}\begin{aligned}\label{at}
\widetilde{a}\ti{12}(\lambda, \mu)&=
\widetilde{r}^+\ti{12}(\lambda,\mu)
-\alpha\ti{12}^\h,\\
\widetilde{d}\ti{12}(\lambda, \mu)&=
\widetilde{r}^-\ti{12}(\lambda,\mu)
+\alpha\ti{12}^\h,
\end{aligned}\end{equation}
and we recall that
\begin{equation}
\widetilde{r}^\pm\ti{12}(\lambda,\mu) = -\frac{1}{4K} \Bigl(\varphi^+(\lambda,\mu) C\ti{12}^\h +
\varphi^-(\lambda,\mu) C\ti{12}^{\alg{a}}
\pm \lambda \pi \rho\ti{12} \delta(\lambda-\mu)
\Bigr).
\end{equation}
The matrices $\widetilde{a}\ti{12}(\lambda, \mu)$ and $\widetilde{d}\ti{12}(\lambda, \mu)$ are skew-symmetric
\begin{equation}
\widetilde{a}\ti{12}(\lambda, \mu) = - \widetilde{a}\ti{21}(\mu,\lambda), \qquad
\widetilde{d}\ti{12}(\lambda, \mu) = - \widetilde{d}\ti{21}(\mu,\lambda).
\end{equation}
Note that these matrices are independent of the asymptotic values $H^\pm$ of $g(x,t)$
defined in equation \eqref{eq31asval}.

In order to extend to any value of $\lambda \in \mathbb{R}$ and $\mu \in \mathbb{R}$, we
introduce the involutive automorphism $\sigma$ of the Lie algebra $\alg{f}$ as
\begin{equation}
\sigma (\g) = \g,\qquad \sigma(\alg{f}^{(1)}) = - \alg{f}^{(1)},
\end{equation}
and its lift $\widehat{\sigma}$ to the Lie group.
We then have the property
\begin{align}
L(-\lambda) &= \sigma(L(\lambda)).
\label{opposite lambda}
\end{align}
This implies that $T_\ell(-\lambda,t)
= \widehat{\sigma}(T_\ell(\lambda,t))$.
Moreover, since $k(\lambda)$ is odd
\beq
\widehat{\Phi}^0(-\lambda,x) = \exp\bigl(-k(\lambda) \Omega x) =
\widehat{\sigma}(\widehat{\Phi}^0(\lambda,x)),
\eeq
hence $\st_\ell (-\lambda,t)
= \widehat{\sigma}(\st_\ell(\lambda,t))$ and
\beq
\st(-\lambda,t) =\widehat{\sigma}(\st(\lambda,t)).
\eeq
This property enables us to extend the Poisson bracket $\{\st(\lambda), \st(\mu)\}$
derived for
$\lambda >0$, $\mu>0$~\eqref{PB stell stell eq1} and \eqref{at} to both the regimes $\lambda <0$, $\mu>0$ and $\lambda<0$, $\mu<0$.

\subsection{The Sine-Gordon case}

To conclude this section, let us comment on the Sine-Gordon case, $F = SO(3)$, $G=SO(2)$, $H=\emptyset$.
The Poisson bracket of the subtracted monodromy matrix is given by~\eqref{PB stell stell eq1} with
\begin{equation}\begin{aligned}\label{atsg}
\widetilde{a}\ti{12}(\lambda, \mu)&=
-\frac{1}{4K} \Bigl(
\varphi^-(\lambda,\mu) C\ti{12}^{\alg{a}}
+ \lambda \pi \rho\ti{12} \delta(\lambda-\mu)
\Bigr), \\
\widetilde{d}\ti{12}(\lambda, \mu)&= -\frac{1}{4K} \Bigl(\varphi^-(\lambda,\mu) C\ti{12}^{\alg{a}}
- \lambda \pi \rho\ti{12} \delta(\lambda-\mu)
\Bigr).
\end{aligned}\end{equation}
This agrees with the result in~\cite{fadtakbook87,Babebook}.
However, it is not obvious that this should be the case since there the Lax matrix with ultralocal Poisson bracket was used, while here we have used the non-ultralocal Lax matrix.
Nevertheless, agreement may be expected since the two Lax matrices are related by a formal gauge transformation~\cite{Vicedo:2017cge}.

Let us first recall that the relation between the non-ultralocal $L_\pm(\lambda)$ and ultralocal $\widetilde{L}_\pm(\lambda)$ Lax connections is
\begin{equation}\label{laxtrans}
\widetilde{L}_\pm(\lambda,x) = \groot(x) L_\pm(\lambda,x) \groot^{-1}(x) - \partial_\pm \groot(x) \groot^{-1}(x),
\end{equation}
where $\groot(x,t) = g^{1/2}(x,t) \in G$.
Since $G$ is abelian, we can make sense of the square root.
The Hamiltonian expressions of the non-ultralocal $L(\lambda)$ and ultralocal $\widetilde{L}(\lambda)$ Lax matrices are
\begin{subequations}\begin{align}
L(\lambda)&= \frac{1}{2K} X + \ha g^{-1} \partial_x g - \ha \lambda m \Omega + \ha \lambda^{-1}
m g^{-1} \Omega g, \label{asymp nul sg}\\
\widetilde{L}(\lambda)&= \frac{1}{2K} X - \ha \lambda m \groot \Omega \groot^{-1} + \ha \lambda^{-1}
m \groot^{-1} \Omega \groot.
\end{align}\end{subequations}
Note that equation \eqref{relgtimeder} becomes $X= K g^{-1} \partial_t g$ in the Sine-Gordon case.
Contrary to $L(\lambda)$, $\widetilde{L}(\lambda)$ does not depend on the spatial derivative of $g$.
This explains why it is ultralocal since the derivative of the Dirac distribution does not appear in its Poisson bracket.

Denoting the generator of $G = SO(2)$ by $T^1$, the asymptotic conditions on $g$ are $g(x,t) \stackrel[x\to \pm\infty]{}{\simeq} \exp(2 \pi Q^\pm T^1) = \mathcal{I}_3$ with $Q^\pm \in \mathbb{Z}$, where the difference $Q^+ - Q^-$ is a physical topological charge since $\pi_1(S^1) = \mathbb{Z}$.
This means that $\groot(x,t) \stackrel[x\to \pm\infty]{}{\simeq} \exp( \pi Q^\pm T^1)$, hence $\lim_{x\to\pm\infty} \groot \Omega \groot^{-1} = \lim_{x\to\pm\infty} \groot^{-1} \Omega \groot = \exp(i\pi Q^\pm) \Omega$.
Therefore, a second difference between the two Lax matrices is that the asymptotic values of $L(\lambda)$ do not depend on $Q^\pm$, however, those of $\widetilde{L}(\lambda)$ do.
This follows since the definition of $\groot$ as a square root introduces an ambiguity. %v2

It follows from the gauge transformation~\eqref{laxtrans} that the transition matrices $T_\ell(\lambda)$ and $\widetilde{T}_\ell(\lambda)$, corresponding to $L(\lambda,x)$ and $\widetilde{L}(\lambda,x)$ respectively, are related as
\begin{equation}\label{montrans}
\widetilde{T}_\ell(\lambda) = \groot(\ell)T_\ell(\lambda) \groot^{-1}(-\ell).
\end{equation}
Since the matrices $U^\pm(x,t)$ are not present for Sine-Gordon, it is clear from the definition~\eqref{2eqresmono} that we can define identical subtracted monodromy matrices starting from the ultralocal or non-ultralocal
Lax matrix by suitably modifying the form of $\widehat{\Phi}^0(\lambda,\pm\ell)$.
Indeed, comparing our subtracted monodromy in the Sine-Gordon case with that in \cite{fadtakbook87,Babebook}, we indeed find that they are related in this way.

\medskip

Actually, in the Sine-Gordon case the equivalence between the ultralocal and non-ultralocal pictures starts even at the level of the Poisson bracket of the transition matrices $T_\ell$
and $\widetilde{T}_\ell$.
In the non-ultralocal case we have the regularised result~\eqref{pb T T continuum}, however, since $\alg{g}$ is one dimensional and $\alpha\ti{12} \in \alg{g}\otimes\alg{g}$ is skew-symmetric, we must have $\alpha\ti{12} = 0$.
This is consistent with $\alpha$ being a solution of the mCYBE on an abelian Lie algebra. %v2
Therefore, we have
\begin{equation}
\{ {T_\ell}\ti{1}(\lambda), {T_\ell}\ti{2}(\mu) \}
=-[r\ti{12}(\lambda,\mu) ,{T_\ell}\ti{1}(\lambda) {T_\ell}\ti{2}(\mu) ],
\end{equation}
where
\begin{equation}\label{526}
r\ti{12}(\lambda,\mu) =
\frac{1}{4K}\bigl( \frac{\mu^2+\lambda^2}{\mu^2-\lambda^2}
C\ti{12}^{(00)} + \frac{2 \lambda \mu}{\mu^2 -\lambda^2} C\ti{12}^{(11)}\bigr).
\end{equation}
Again, this result agrees with that found from the ultralocal Lax matrix~\cite{fadtakbook87,Babebook}.

To see why this happens, let us rewrite equations~\eqref{laxtrans} and \eqref{montrans} as
\begin{equation}\begin{aligned}
L(\lambda,x) & = \groot^{-1}(x) \widetilde{L}(\lambda,x) \groot(x) +
\groot^{-1}(x)\partial_x \groot(x),
\\
T_\ell(\lambda) & = \groot^{-1}(\ell) \widetilde{T}_\ell(\lambda) \groot(-\ell) .
\end{aligned}\end{equation}
A first computation shows that
\begin{equation} \label{pbneededgt}
\{ \widetilde{L}\ti{1}(\lambda,x), \groot\ti{2}(y) \}
= \{L\ti{1}(\lambda,x), \groot\ti{2}(y) \} = \groot\ti{2}(x) s\ti{12} \delta(x-y)
= s\ti{12}\groot\ti{2}(x) \delta(x-y),
\end{equation}
where we recall $s\ti{12} = 1/(4K) \, C\ti{12}^{(00)}$, hence $s\ti{12}$ and $\groot$ commute since $\alg{g}$ is abelian.
From this, we can compute
\begin{equation}
\{ \overleftarrow{U}\ti{1}(x_1,x_2,-\widetilde{L}(\lambda)), \groot\ti{2}(y) \}
= - \overleftarrow{U}\ti{1}(y,x_2,-\widetilde{L}(\lambda)) s\ti{12}
\groot\ti{2}(y) \overleftarrow{U}\ti{1}(x_1,y,-\widetilde{L}(\lambda))
\chi(y;[x_1,x_2]),
\end{equation}
where $\chi(y;[x_1,x_2])=1$ if $y \in [x_1,x_2]$ and zero otherwise.
Taking $y$ to be coincident with $x_1$ and $x_2$, this implies
\begin{equation}\begin{split}\label{uh1}
\{ \overleftarrow{U}\ti{1}(x_1,x_2,-\widetilde{L}(\lambda)), \groot\ti{2}(x_1) \}
& = - \overleftarrow{U}\ti{1}(x_1,x_2,-\widetilde{L}(\lambda)) s\ti{12}
\groot\ti{2}(x_1),
\\
\{ \overleftarrow{U}\ti{1}(x_1,x_2,-\widetilde{L}(\lambda)), \groot\ti{2}(x_2) \}
&= -
\groot\ti{2}(x_2) s\ti{12} \overleftarrow{U}\ti{1}(x_1,x_2,-\widetilde{L}(\lambda)),
\end{split}\end{equation}
and
\begin{equation}\begin{split}\label{uh2}
\{ \overleftarrow{U}\ti{1}(x_1,x_2,-\widetilde{L}(\lambda)), \groot^{-1}\ti{2}(x_1) \}
& = \overleftarrow{U}\ti{1}(x_1,x_2,-\widetilde{L}(\lambda)) s\ti{12}
\groot^{-1}\ti{2}(x_1),
\\
\{ \overleftarrow{U}\ti{1}(x_1,x_2,-\widetilde{L}(\lambda)), \groot^{-1}\ti{2}(x_2) \}
&=
\groot^{-1}\ti{2}(x_2) s\ti{12} \overleftarrow{U}\ti{1}(x_1,x_2,-\widetilde{L}(\lambda)),
\end{split}\end{equation}
again using that $s\ti{12}$ and $\groot$ commute.
It follows that the Poisson bracket of two non-ultralocal transition matrices with coincident endpoints is given by
\begin{equation}\begin{aligned}
&\{ \overleftarrow{U}\ti{1}(x_1,x_2,-L(\lambda)), \overleftarrow{U}\ti{2}(x_1,x_2,-L(\mu)) \}
\\=
\,&\{\groot^{-1}\ti{1}(x_2)\overleftarrow{U}\ti{1}(x_1,x_2,-\widetilde{L}(\lambda))\groot\ti{1}(x_1),
\groot^{-1}\ti{2}(x_2)\overleftarrow{U}\ti{2}(x_1,x_2,-\widetilde{L}(\mu))\groot\ti{2}(x_1)\}
\\=
\,&\groot^{-1}\ti{1}(x_2)\groot^{-1}\ti{2}(x_2)\{\overleftarrow{U}\ti{1}(x_1,x_2,-\widetilde{L}(\lambda)),
\overleftarrow{U}\ti{2}(x_1,x_2,-\widetilde{L}(\mu))\}\groot\ti{1}(x_1)\groot\ti{2}(x_1),
\end{aligned}\end{equation}
where we use equations~\eqref{uh1} and \eqref{uh2}, and that $s\ti{12}$ is symmetric.
Finally, taking $x_1 = -\ell$ and $x_2 = \ell$, we find
\begin{equation}\begin{split}
\{T_\ell{}\ti{1}(\lambda),T_\ell{}\ti{2}(\mu)\} &=
\groot^{-1}\ti{1}(\ell)\groot^{-1}\ti{2}(\ell)
\{\widetilde{T}_\ell{}\ti{1}(\lambda),\widetilde{T}_\ell{}\ti{2}(\mu)\}
\groot\ti{1}(-\ell)\groot\ti{2}(-\ell)
\\ & =
-\groot^{-1}\ti{1}(\ell)\groot^{-1}\ti{2}(\ell)
[r\ti{12}(\lambda,\mu) ,\widetilde{T}_\ell{}\ti{1}(\lambda) \widetilde{T}_\ell{}\ti{2}(\mu) ]
\groot\ti{1}(-\ell)\groot\ti{2}(-\ell)
\\ & =
-[r\ti{12}(\lambda,\mu) ,T_\ell{}\ti{1}(\lambda) T_\ell{}\ti{2}(\mu) ],
\end{split}\end{equation}
where we have used that
\begin{equation} \label{rtransformnul}
\groot^{-1}\ti{1}(x)\groot^{-1}\ti{2}(x) r\ti{12}(\lambda,\mu) \groot\ti{1}(x)\groot\ti{2}(x) =r\ti{12}(\lambda,\mu),
\end{equation}
since $r\ti{12}$~\eqref{526} is a linear combination of Casimirs of $\alg{g}$ and $\alg{f}$.
Thus we find that the Poisson brackets of the ultralocal and non-ultralocal transition matrices $\widetilde{T}_\ell$ and $T_\ell$ coincide in the Sine-Gordon case as claimed.

Let us note that these results also follow from those in section 6 of \cite{Delduc:2015xdm},
where the effect of a gauge transformation on an $r/s$ Poisson bracket has been worked out. Performing the formal gauge transformation that goes from the ultralocal Lax matrix to the non-ultralocal Lax matrix, equations \eqref{pbneededgt} and \eqref{rtransformnul} then imply that the matrix $r$ and $s=0$ are transformed into $r$ itself and $s\ti{12}=(1/4K)C^{(00)}\ti{12}$.

\section{Analysis of the Poisson bracket of the subtracted monodromy}\label{sec6}

Before taking the limit $\ell \to \infty$, we have the bracket \eqref{pb T T continuum} for transition matrices. This bracket satisfies the Jacobi identity when $\xi^2 = 1/(16 K^2)$. In this section, we first prove that the Jacobi identity remains valid after taking the limit of an infinite interval.
We then show that it remains possible to construct an infinite number of conserved quantities in involution taking $\xi$ to be non-zero.

\subsection{Jacobi identity}\label{jacobi}

As usual for a quadratic algebra of the form \eqref{PB stell stell eq1} with $\widetilde{a}$ and
$\widetilde{d}$ skew-symmetric, we have
\beqz
\{ \{{\st}\ti{1}(\lambda_1) , {\st}\ti{2}(\lambda_2) \}, {\st}\ti{3}(\lambda_3) \}
+ \mathrm{c.p.} = Y\ti{123}(\widetilde{a}) {\st}\ti{1}(\lambda_1) {\st}\ti{2}(\lambda_2)
{\st}\ti{3}(\lambda_3) - {\st}\ti{1}(\lambda_1) {\st}\ti{2}(\lambda_2)
{\st}\ti{3}(\lambda_3) Y\ti{123}(\widetilde{d}) ,
\eeqz
where $\mathrm{c.p.}$ stands for cyclic permutations and where $Y\ti{123}(r)$ represents the l.h.s. of the CYBE
\beq \label{Y r}
Y\ti{123}(r) = [r\ti{12},
r\ti{13}]+ [r\ti{12},
r\ti{23}] + [r\ti{13},
r\ti{23}], \qquad r\ti{12}(\lambda_1,\lambda_2) = - r\ti{21}(\lambda_2,\lambda_1).
\eeq
The dependence in spectral parameters is left implicit.
The matrices $\widetilde{a}$ and $\widetilde{d}$ are obtained by summing terms proportional to $C^\h$, $\rho$, $\alpha^\h$ and $C^{\alg{a}}$.
Noting that $Y\ti{123}(C^{\alg{a}}) = 0$ since $\alg{a}$ is abelian, we shall proceed by determining
$Y\ti{123}(\rho)$, $Y\ti{123}(C^{\alg{h}})$ and $Y\ti{123}(\alpha^{\alg{h}})$.
We will then argue that the cross-terms vanish, allowing us to demonstrate that both $\widetilde{a}$ and $\widetilde{d}$ satisfy the same mCYBE, implying that the Jacobi identity is satisfied.

\paragraph{Equation satisfied by $\rho=\ad_\Omega$.}
Let us prove that
\beq \label{sortofcyberho}
Y\ti{123}(\rho)
=[C\ti{12},C\ti{13}] -[C^\h\ti{12},C^\h\ti{13}].
\eeq
or equivalently in terms of operators
\beq \label{eqrho}
[\rho \alga,\rho \algb]
-\rho (
[\rho \alga, \algb] + [\alga, \rho \algb]
)
-[\alga,\algb]
+[\alga^\h,\algb^\h] =0, \qquad \alga,\algb\in\f,
\eeq
where for any $\alga\in \f$, $\alga^\h$ denotes its projection onto $\h$.
To do so, we consider $\algb$ and $\algb$ restricted to all possible subspaces in the decomposition~\eqref{haompommd}.
Summarising, we find
\beqz
\begin{array}{||c|c||c|c|c|c|c||}
\hline
\alga \in \ & \algb\in \ &[\rho \alga,\rho \algb] = \ &-\rho
[\rho \alga, \algb]= \ &-\rho[\alga, \rho \algb] = \ &-[\alga,\algb] = \ &[\alga^\h,\algb^\h] = \ \\
\hline
\h^\mathbb{C} &\h^\mathbb{C}&0&0&0&-[\alga,\algb]&+[\alga,\algb]\\
\h^\mathbb{C}&\alg{a}^\mathbb{C}&0&0&0&0&0\\
\h^\mathbb{C}&\omega^\pm &0&0&+[\alga,\algb]&-[\alga,\algb]&0\\
\alg{a}^\mathbb{C} &\alg{a}^\mathbb{C}&0&0&0&0&0\\
\alg{a}^\mathbb{C} & \omega^\pm&0&0&+[\alga,\algb]&-[\alga,\algb]&0\\
\omega^\pm&\omega^\pm&0&0&0&0&0\\
\omega^+&\omega^-&+[\alga,\algb]&0&0&-[\alga,\algb]&0\\
\hline
\end{array}
\eeqz
where we have used table~\eqref{list com basis adomega}.
It is then immediate that the sum for each line vanishes and
the identity~\eqref{eqrho} is satisfied.

\paragraph{Equation satisfied by $\varphi^+(\lambda,\mu) C\ti{12}^\h$.}
Let us introduce
\beq
\chi^\h\ti{12}(\lambda,\mu) = \varphi^+(\lambda,\mu) C\ti{12}^\h,
\eeq
where $\varphi^+$, defined in \eqref{def phipm}, may be written as
\beq
\varphi^+(\lambda,\mu) = - \PP \frac{\mu}{\lambda-\mu} - \frac{\lambda}{\lambda+\mu}
= - \PP \frac{\lambda}{\lambda-\mu} + \frac{\mu}{\lambda+\mu}.
\eeq
Using this freedom and the property \eqref{property Casimir} for the Casimir $C\ti{12}^\h$, we obtain
\beq
Y\ti{123}(\chi^\h) = z(\lambda_1,\lambda_2,\lambda_3) [C^\h\ti{12},C^\h\ti{13}],
\eeq
where
\begin{align}
z(\lambda_1,\lambda_2,\lambda_3) &= \bigl(-\PP \frac{\lambda_2}{\lambda_1-\lambda_2} - \frac{\lambda_1}{\lambda_1+\lambda_2}
\bigr)
\bigl(
-\PP \frac{\lambda_1}{\lambda_1-\lambda_3} + \frac{\lambda_3}{\lambda_1+\lambda_3}
\bigr)
\nonumber\\
&\qquad - \bigl(
-\PP \frac{\lambda_1}{\lambda_1-\lambda_2} + \frac{\lambda_2}{\lambda_1+\lambda_2}
\bigr)
\bigl(
-\PP \frac{\lambda_2}{\lambda_2-\lambda_3} + \frac{\lambda_3}{\lambda_2+\lambda_3}
\bigr)
\\\nonumber
&\qquad +\bigl(
- \PP\frac{\lambda_1}{\lambda_1-\lambda_3}+ \frac{\lambda_3}{\lambda_1+\lambda_3}
\bigr)
\bigl(
-\PP \frac{\lambda_2}{\lambda_2-\lambda_3} + \frac{\lambda_3}{\lambda_2+\lambda_3}
\bigr).
\end{align}
The quantity $z(\lambda_1,\lambda_2,\lambda_3)$ contains terms with zero, one or two
principal values. Each of the three pairs of terms with a single principal value
is actually non singular.
Their sum gives
\beq
\lambda_3 \frac{\lambda_1+\lambda_2+\lambda_3}{(\lambda_1+\lambda_3)(\lambda_2+\lambda_3)}
+ \frac{\lambda_1 \lambda_2}{(\lambda_1+\lambda_2)(\lambda_2+\lambda_3)}
+\frac{\lambda_1 \lambda_2}{(\lambda_1+\lambda_2)(\lambda_1+\lambda_3)},
\eeq
and the sum of this term with those that have no principal value gives $1$.

The sum of the terms containing two principal values is equal to
\beq\label{eq:omega}
\omega(\lambda_1,\lambda_2,\lambda_3) =
\PP\frac{\lambda_2}{\lambda_1-\lambda_2}
\PP\frac{\lambda_1}{\lambda_1-\lambda_3}
- \PP\frac{\lambda_1}{\lambda_1-\lambda_2}
\PP\frac{\lambda_2}{\lambda_2-\lambda_3}
+ \PP\frac{\lambda_1}{\lambda_1-\lambda_3}
\PP\frac{\lambda_2}{\lambda_2-\lambda_3}.
\eeq
At this point, we use one of the known methods showing that $\PP\frac{1}{\mu-\lambda}C\ti{12}$ is a skew-symmetric solution of the mCYBE on the loop algebra ${\cal L}\f$.
One such method proves that this holds in terms of operators.
Indeed, up to subtleties (see, for instance, \cite{Vicedo:2010qd}), $\PP\frac{1}{\mu-\lambda}C\ti{12}$ is the kernel of the difference of the projectors onto the regular and pole parts of ${\cal L}\f$.
Such an operator corresponds to the Adler-Kostant-Symes construction \cite{Adler:1979ib,Kostant:1979qu,Symes:1981}, hence is a solution of the mCYBE on the loop algebra of $\f$.
A second method, which we will use here, involves working in terms of the kernels and making use of the Poincar\'e-Bertrand formula (see, for instance, \cite{SemenovTianShansky:1983ik}).
This formula is related to the Hilbert transform.

The Hilbert transform $H f$ of a function $f$ is defined by
\beq \label{def hilbert transform}
H f(x) = \frac{1}{\pi} \PP \int_{-\infty}^{+\infty} dt \frac{f(t)}{t-x}.
\eeq
It satisfies the property (see, for instance, the last corollary of \cite{erlove77})
\beq \label{corollary hilbert}
Hf Hg - H \bigl( f Hg + g H f\bigr) = f g.
\eeq
Note the similarity between this equation and the mCYBE.
For a function $f$, we define $\widehat{f}(\lambda) = \lambda f(\lambda)$.
In terms of $\widehat{f}$ and $\widehat{g}$, the property \eqref{corollary hilbert}
is written
\beq \label{corollary hilbert 2}
H\widehat{f}(\lambda_3) H\widehat{g}(\lambda_3) - H \bigl( \widehat{f} H \widehat{g} + \widehat{g} H \widehat{f}\bigr)(\lambda_3)
= \lambda_3^2 f(\lambda_3) g(\lambda_3).
\eeq

Now consider the integral
\beq
\int d\lambda_1 d\lambda_2 \; {\omega}(\lambda_1,\lambda_2,\lambda_3)
f(\lambda_1) g(\lambda_2),
\eeq
where $\omega$ is defined in~\eqref{eq:omega}.
It is the sum of the following three terms
\begin{equation}\begin{aligned}
\PP\int\frac{d\lambda_1}{\lambda_1-\lambda_3} \lambda_1 f(\lambda_1) \PP\int\frac{d\lambda_2}{\lambda_1-\lambda_2} \lambda_2 g(\lambda_2)
&= - \pi^2 H(\widehat{f} H \widehat{g})(\lambda_3),\\
- \PP\int\frac{d\lambda_2}{\lambda_2-\lambda_3} \lambda_2 g(\lambda_2) \PP\int\frac{d\lambda_1}{\lambda_1-\lambda_2}
\lambda_1 f(\lambda_1)
&= - \pi^2 H(\widehat{g} H \widehat{f})(\lambda_3),\\
\PP\int\frac{d\lambda_1}{\lambda_1-\lambda_3} \lambda_1 f(\lambda_1)
\PP\int \frac{d\lambda_2}{\lambda_2-\lambda_3} \lambda_2 g(\lambda_2)
&= \pi^2 H\widehat{f}(\lambda_3) H\widehat{g}(\lambda_3),
\end{aligned}\end{equation}
hence, by the property~\eqref{corollary hilbert 2} is equal to
$\pi^2 \lambda_3^2 f(\lambda_3) g(\lambda_3)$.
Therefore,
${\omega}(\lambda_1,\lambda_2,\lambda_3) = \pi^2 \lambda_3^2 \delta(\lambda_1-\lambda_2)
\delta(\lambda_1-\lambda_3)$, $z(\lambda_1,\lambda_2,\lambda_3) = 1+ \pi^2 \lambda_3^2 \delta(\lambda_1-\lambda_2)
\delta(\lambda_1-\lambda_3)$ and we find that
\begin{equation}
Y\ti{123}(\chi^\h) =\Bigl(1 + \pi^2 \lambda_1^2
\delta(\lambda_1-\lambda_2) \delta(\lambda_1
-\lambda_3)\Bigr)
[C^\h\ti{12}, C^\h\ti{13}]. \label{eq for chi}
\end{equation}

\paragraph{Equation satisfied by $\alpha^\h\ti{12}$.}

By definition, $\alpha^\h$ satisfies the equation
\beq \label{repeat eq alphah}
Y\ti{123}(\alpha^\h)
= -\xi^2 [C^\h\ti{12},C^\h\ti{13}],
\eeq
where, for now, we leave the value of $\xi$ free.

\paragraph{Cross-terms.}
There are six classes of cross-terms.
It is immediate that the cross-terms between $\chi^{\alg{h}}$ and $C^{\alg{a}}$ and between $\alpha^{\alg{h}}$ and $C^{\alg{a}}$ vanish since $[\alg{h},\alg{a}] = 0$.

For the cross-terms between $\beta = -\frac{1}{4K}\varphi^+ C^{\alg{h}} -\frac{1}{4K}\varphi^- C^{\alg{a}} \mp \alpha^{\alg{h}}$ and $\rho$, we group terms proportional to the same Dirac distribution.
For instance, those proportional to
$\delta(\lambda_1-\lambda_3)$ are
\begin{equation}\begin{split}
& -\frac{\pi\lambda_1\delta(\lambda_1-\lambda_3)}{4K} \Bigl(
[\beta{}\ti{12}(\lambda_1,\lambda_2),
\rho\ti{13}] +
[ \rho\ti{13}, \beta{}\ti{23}(\lambda_2,\lambda_3)]
\Bigr)
\\ & \qquad \qquad \qquad
=\frac{\pi \lambda_1 \delta(\lambda_1-\lambda_3)}{4K}
\Bigl([ \rho\ti{13},\beta\ti{12}(\lambda_1, \lambda_2) + \beta\ti{32}(\lambda_1,\lambda_2)]\Bigr),
\end{split}\end{equation}
where we have used the Dirac distribution to set $\lambda_3 = \lambda_1$.
Now we can use
\begin{equation}\begin{split}\label{eq:ch}
[ \rho\ti{13}, \beta\ti{12}(\lambda_1, \lambda_2) + \beta\ti{32}(\lambda_1,\lambda_2)]
& =
[ [\Omega\ti{1},C\ti{13}],\beta\ti{12}(\lambda_1, \lambda_2) + \beta\ti{32}(\lambda_1,\lambda_2)]
\\ & =
[\Omega\ti{1} ,[ C\ti{13},\beta\ti{12}(\lambda_1, \lambda_2) + \beta\ti{32}(\lambda_1,\lambda_2)] ]= 0,
\end{split}\end{equation}
where we have used that $[\Omega\ti{1},\beta\ti{12}] = 0$ since $\beta \in \alg{h} \otimes \alg{h} + \alg{a} \otimes \alg{a}$.
It follows that the terms proportional to $\delta(\lambda_1-\lambda_3)$ vanish, and, by cyclicity, so do all cross-terms of this type.

Finally, for the cross-terms between $\chi^{\alg{h}}$ and $\alpha^{\alg{h}}$, we group the terms as
\begin{equation}
\mp\frac{1}{4K}\big([\chi^{\alg{h}}\ti{12},\alpha^{\alg{h}}\ti{13}+\alpha^{\alg{h}}\ti{23}]\big)
\mp\frac{1}{4K}\big([\chi^{\alg{h}}\ti{13},\alpha^{\alg{h}}\ti{21}+\alpha^{\alg{h}}\ti{23}]\big)
\mp\frac{1}{4K}\big([\chi^{\alg{h}}\ti{23},\alpha^{\alg{h}}\ti{21}+\alpha^{\alg{h}}\ti{31}]\big).
\end{equation}
Since $\chi^{\alg{h}} \propto C^{\alg{h}}$ and $\alpha^{\alg{h}}$ is independent of the spectral parameters, it follows that these terms vanish due to the property \eqref{property Casimir} for the Casimir $C\ti{12}^\h$.

\paragraph{Equation satisfied by $\widetilde{a}$ and $\widetilde{d}$.}
Putting the above results together, we find that
\begin{equation}\begin{aligned}\label{Yfinal}
Y\ti{123}(\widetilde{a}) &=
\frac{1}{16 K^2}
\Bigl(1 + \pi^2 \lambda_1^2
\delta(\lambda_1-\lambda_2) \delta(\lambda_1
-\lambda_3)\Bigr)
[C^\h\ti{12}, C^\h\ti{13}] \\
&\quad +\frac{1}{16 K^2}
\pi^2 \lambda_1^2
\delta(\lambda_1-\lambda_2) \delta(\lambda_1
-\lambda_3) \Bigl(
[C\ti{12},C\ti{13}] -[C^\h\ti{12},C^\h\ti{13}]
\Bigr) - \xi^2 [C^\h\ti{12},C^\h\ti{13}] \\
&= (\frac{1}{16 K^2} - \xi^2) [C^\h\ti{12},C^\h\ti{13}]
+\frac{1}{16 K^2}
\pi^2 \lambda_1^2
\delta(\lambda_1-\lambda_2) \delta(\lambda_1
-\lambda_3)
[C\ti{12},C\ti{13}].
\end{aligned}\end{equation}
Setting $\xi^2 = 1/(16 K^2)$, the matrix $\widetilde{a}$ satisfies the mCYBE on the loop algebra
${\cal L}\f$
\begin{equation}\begin{aligned}
[\widetilde{a}\ti{12}(\lambda_1, \lambda_2),
\widetilde{a}\ti{13}(\lambda_1, \lambda_3)]
+
[\widetilde{a}\ti{12}(\lambda_1, \lambda_2),
\widetilde{a}\ti{23}(\lambda_2, \lambda_3)]
+
[\widetilde{a}\ti{13}(\lambda_1, \lambda_3),
\widetilde{a}\ti{23}(\lambda_2, \lambda_3)] \qquad \qquad \\
=\frac{1}{16 K^2}
\pi^2 \lambda_1^2
\delta(\lambda_1-\lambda_2) \delta(\lambda_1
-\lambda_3)
[C\ti{12},C\ti{13}].
\end{aligned}\end{equation}
The matrix $\widetilde{d}$ satisfies the same equation.
This ensures that the
Poisson bracket \eqref{PB stell stell eq1} of the subtracted monodromy satisfies the Jacobi identity.

\subsection{Conserved quantities in involution}\label{sec:conserved}

To construct conserved quantities in involution, we start by rewriting the Poisson bracket \eqref{PB stell stell eq1} as
\beq
\{ {\st}\ti{1}(\lambda) , {\st}\ti{2}(\mu) \} =
[\widetilde{d}\ti{12}(\lambda, \mu) , {\st}\ti{1}(\lambda) {\st}\ti{2}(\mu) ]
+\bigl(\widetilde{a}\ti{12}(\lambda,\mu) - \widetilde{d}\ti{12}(\lambda,\mu)\bigr)
{\st}\ti{1}(\lambda) {\st}\ti{2}(\mu).
\eeq
The difference between the matrices $\widetilde{a}$ and $\widetilde{d}$ is equal to
\beq
- \frac{\lambda \pi}{2 K} \rho\ti{12} \delta(\lambda-\mu) - 2 \alpha^\h\ti{12}.
\eeq
Since the matrix $\rho\ti{12}$ is skew-symmetric and is multiplied by the Dirac distribution, it does not contribute to the Poisson bracket of $\tr\st(\lambda)$ with $\tr\st(\mu)$.
As a result, we have
\beq
\{ \tr \st(\lambda), \tr \st(\mu) \} = - 2 \tr\ti{1}\tr\ti{2} \bigl(
\alpha\ti{12}^\h \st\ti{1}(\lambda)\st\ti{2}(\mu)
\bigr).
\eeq
It follows that the conserved quantities $\tr\st(\lambda)$ are not in involution.
However, recalling the discussion in subsection \ref{subsec: time evolution}, the quantity
$\tr\st_\gamma(\lambda)$, with $\st_\gamma(\lambda)$ defined in equation~\eqref{stgamma},
\beqz
\st_\gamma (\lambda,t) = (\gamma^+(\lambda))^{-1} \st(\lambda,t) \gamma^-(\lambda)
\eeqz
is also conserved.
By the cyclicity of the trace, we have
\beq
\tr\st_\gamma(\lambda) = \tr \bigl(
\gamma(\lambda) \st(\lambda)
\bigr), \qquad \gamma(\lambda) = \gamma^-(\lambda) (\gamma^+(\lambda))^{-1},
\eeq
and we also recall that $\gamma^\pm$, and hence $\gamma$, should commute with $\Omega$.
Following \cite{Freidel:1991jx,Freidel:1991jv}, since
\begin{equation}\begin{split}
\{ {\st_\gamma}\ti{1}(\lambda), {\st_\gamma}\ti{2}(\mu)\} = & \,
(\gamma^+\ti{1}(\lambda))^{-1} (\gamma^+\ti{2}(\mu))^{-1}
\widetilde{a}\ti{12}(\lambda,\mu) \gamma^+\ti{1}(\lambda) \gamma^+\ti{2}(\mu)
{\st_\gamma}\ti{1}(\lambda) {\st_\gamma}\ti{2}(\mu) \\ & \,
-
{\st_\gamma}\ti{1}(\lambda) {\st_\gamma}\ti{2}(\mu)
(\gamma^-\ti{1}(\lambda))^{-1} (\gamma^-\ti{2}(\mu))^{-1}
\widetilde{d}\ti{12}(\lambda,\mu)
\gamma^-\ti{1}(\lambda) \gamma^-\ti{2}(\mu),
\end{split}\end{equation}
we look for $\gamma^\pm(\lambda)$ such that the Poisson bracket above becomes a commutator,
\begin{equation}
\{ {\st_\gamma}\ti{1}(\lambda), {\st_\gamma}\ti{2}(\mu)\} = \,
[(\gamma^+\ti{1}(\lambda))^{-1} (\gamma^+\ti{2}(\mu))^{-1}
\widetilde{a}\ti{12}(\lambda,\mu) \gamma^+\ti{1}(\lambda) \gamma^+\ti{2}(\mu),
{\st_\gamma}\ti{1}(\lambda) {\st_\gamma}\ti{2}(\mu) ].
\end{equation}
In addition to commuting with $\Omega$, this means that $\gamma(\lambda)$ should satisfy
\begin{equation}
\gamma\ti{1}(\lambda)\gamma\ti{2}(\mu)
\widetilde{a}\ti{12} (\gamma\ti{1}(\lambda))^{-1}(\gamma\ti{2}(\mu))^{-1}
= \widetilde{d}\ti{12}. \label{condatdt}
\end{equation}
For simplicity, we may assume that $\gamma(\lambda)$ is independent of the spectral parameter, in which case~\eqref{condatdt} becomes
\begin{subequations}
\begin{align}
(\gamma\ti{1})^{-1}( \gamma\ti{2})^{-1}
C^\h\ti{12} \gamma\ti{1} \gamma\ti{2}
&= C^\h\ti{12}, \label{cond gamma Ch}\\
(\gamma\ti{1})^{-1} (\gamma\ti{2})^{-1} C\ti{12}^{\alg{a}}
\gamma\ti{1} \gamma\ti{2}
&=C\ti{12}^{\alg{a}}, \label{cond gamma Om} \\
(\gamma\ti{1})^{-1} (\gamma\ti{2})^{-1}
\alpha^\h\ti{12} \gamma\ti{1}\gamma\ti{2}
& =-\alpha^\h\ti{12}, \label{cond gamma alpha}
\\
(\gamma\ti{1})^{-1} (\gamma\ti{2})^{-1}
\rho\ti{12} \gamma\ti{1} \gamma\ti{2}
&= -\rho\ti{12}. \label{gamma on rho}
\end{align}
\end{subequations}
Since $\gamma$ commutes with $\Omega$ and $\rho\ti{12}$ is the kernel of the adjoint action of $\Omega$, the condition \eqref{gamma on rho} cannot be satisfied.
However, as for $\tr\st$, the matrix $\rho\ti{12}$ does not contribute to the Poisson bracket of $\tr\st_{\gamma}$ with $\tr\st_{\gamma}$, hence we can drop this equation.
For the spheres, $F=SO(N+1)$, $G=SO(N)$, $H=SO(N-1)$, in appendix \ref{spheres} we construct a quantity $\gamma$, which is independent of the spectral parameter, commutes with $\Omega$ and is also solution of equations \eqref{cond gamma Ch}, \eqref{cond gamma Om} and \eqref{cond gamma alpha}.
We conclude that the associated quantities
$\tr \st_\gamma(\lambda)$ are both conserved and in involution.

\medskip

We have shown that going from $\st(\lambda)$ to $\st_\gamma(\lambda)$ has the effect of changing the trace and the Poisson bracket such that the Poisson bracket of $\st_\gamma(\lambda)$ is a commutator, hence $\tr \st_\gamma(\lambda)$ are in involution.
In subsection \ref{subsec: time evolution}, we also introduced in equation \eqref{tauomega} a third subtracted monodromy
\beqz
\st_\Omega(\lambda,t) = e^{-t \widetilde{k}(\lambda)\Omega} \st(\lambda) e^{t \widetilde{k}(\lambda)\Omega}. %v2
\eeqz
Recall that $\st_\Omega(\lambda,t)$ has the property of being conserved and has the same trace as $\st(\lambda)$.
Moreover, it also has the same Poisson bracket \eqref{PB stell stell eq1} as $\st(\lambda)$.
This is a consequence of $\widetilde{a}\ti{12}(\lambda,\mu)$ and $\widetilde{d}\ti{12}(\lambda,\mu)$ being invariant under the adjoint action of $\exp(\mp({\widetilde{k}}(\lambda) \Omega\ti{1} +{\widetilde{k}}(\mu) \Omega\ti{2}) t)$.
It is clear that $C^\h\ti{12}$, $C\ti{12}^{\alg{a}}$ and $\alpha^\h\ti{12}$ are invariant under this action.
For the terms proportional to $\rho\ti{12} \delta(\lambda-\mu)$, the invariance follows since $\rho\ti{12}$ is the kernel of the adjoint action of $\Omega$.

Of course, starting from the conserved subtracted monodromy $\st_\Omega(\lambda,t)$ and its Poisson bracket, we can define $\st_{\Omega,\gamma}(\lambda,t) = (\gamma^+(\lambda))^{-1} \st_\Omega(\lambda,t) \gamma^-(\lambda)$.
It is clear that $\st_{\Omega,\gamma}(\lambda,t)$ leads to the same conserved quantities in involution as $\st_{\gamma}(\lambda)$ since $\tr\st_{\Omega,\gamma}(\lambda,t) =\tr\st_{\gamma}(\lambda)$. %v2

\medskip

Finally, let us note the important open questions of the completeness and independence of the conserved charges in involution.
While we have constructed an infinite set of conserved quantities in involution, it is unclear if this can be extended, e.g., by taking a more general choice of $\gamma$ depending on the spectral parameter.
It would also be interesting to study more explicitly the form of the conserved charges that we find.
In particular, it has been shown that the Hamiltonian can be extracted from the subtracted monodromy $\st_\Omega(\lambda,t)$ in~\cite{Hollowood:2010dt}.
However, it is not clear if it is in the set of conserved quantities in involution that we find. %v2

\section{Conclusion}\label{secconc}

In this article, we have completed the first stages in the computation of the Poisson bracket of the subtracted monodromy matrix of SSSG theories defined on an infinite interval.
In order to do so, due to the non-ultralocality, we regularised the ill-defined Poisson bracket of the transition matrices, introducing a parameter $\xi$.
$\xi = 0$, i.e. $\alpha^{\alg{h}} = 0$, corresponds to the `standard' regularisation \cite{Maillet:1985fn,Maillet:1985ek}, while $\xi^2 = 1/(16 K^2)$ is the unique value for which the Poisson bracket satisfies the Jacobi identity, as shown in subsection~\ref{jacobi}.
Let us note that for the Sine-Gordon and Complex Sine-Gordon models, for which $H$ is empty and abelian respectively, $\alpha^{\alg{h}} = 0$ identically and the Jacobi identity is satisfied for any $\xi$.
However, this is no longer true for non-abelian $H$.
In subsection~\ref{sec:conserved}, we demonstrated that for non-zero $\xi$ it remains possible to construct an infinite number of conserved quantities in involution by introducing the matrix $\gamma$ in the definition of the subtracted monodromy.

In order to complete the analysis, two further stages need to be taken into account.
The most important is related to the non-trivial asymptotic behavior of the fields.
This has implications for the correct definition of the WZ term in the action \cite{Hollowood:2013oca}.
In turn, amending the action \eqref{action sssg} modifies the canonical analysis.
The modified symplectic form is known for the WZW and gauged WZW models with boundaries \cite{Gawedzki:2001ye,Gawedzki:2001rm} together with their corresponding boundary conditions \cite{Alekseev:1998mc,Elitzur:2001qd}.
However, to our knowledge, the corresponding Poisson brackets have not yet been written down.
Furthermore, the setup we are considering is subtly different, since we are interested in the asymptotic behavior of configurations rather than boundary conditions for a finite interval.
It is therefore a key step to determine the Poisson brackets of the fields $U^\pm(x,t)$ defined for $x\to\pm \infty$.
These fields appear in the asymptotic behavior \eqref{asymptotic sssg} of $g$ and $A_\mu$ and in the definition \eqref{2eqresmono} of the restricted monodromy matrix.

The final step is to address the reality conditions and the matrix $\alpha$.
For $\lambda$ real, the Lax matrix satisfies $L^\dagger(\lambda) = - L(\lambda)$.
This implies a consistency condition at the level of the Poisson bracket \eqref{pbLLcheck} of the Lax matrix.
This is ensured by the properties $(r\ti{12}(\lambda,\mu))^\dagger = r\ti{12}(\lambda,\mu)$ and similarly for $s\ti{12}(\lambda,\mu)$ for $\lambda$ and $\mu$ real.
However, the matrix $\alpha\ti{12}$ does not satisfy $(\alpha\ti{12})^\dagger = \alpha\ti{12}$.
This is related to the fact that there is no split skew-symmetric solution of mCYBE on a compact Lie algebra.
This hinders a proper use of the `lattice' regularisation consistent with the property $\st^\dagger(\lambda) = \st^{-1}(\lambda)$ for $\lambda$ real.
The extent to which this is a problem remains to be understood.

\medskip

A curious feature of the results obtained is the similarity with the tree-level S-matrix for SSSG theories.
If we take $\lambda \neq \mu$ and use the `standard' regularisation $\alpha^{\alg{h}} = 0$, then $\widetilde{a}\ti{12} = \widetilde{d}\ti{12}$ in~\eqref{at}.
Setting $\lambda/\mu = e^{\theta}$, we find that they are proportional to $\operatorname{coth} \theta C\ti{12}^{\mathfrak{h}} - \operatorname{csch} \theta C\ti{12}^{\mathfrak{a}}$.
Computing the tree-level S-matrix for perturbative excitations as in \cite{Hoare:2010fb} one finds $S\ti{12} \propto\operatorname{coth} \theta C^{\alg{h}}\ti{12} - \operatorname{csch}\theta \id\ti{12}$, where $\theta$ is the difference of rapidities and generators are evaluated in the vector representation of $\alg{h} = \mathfrak{so}(N-1)$.
The CYBE is not satisfied and is non-vanishing by a term proportional to the structure constants of the Lie algebra $\h$.
Note that, for the Sine-Gordon and Complex Sine-Gordon theories, this means that the CYBE is satisfied.
This is analogous to what happens for the `standard' regularisation of the Poisson bracket of the subtracted monodromy, for which the Jacobi identity is also not satisfied by terms proportional to the structure constants of $\h$, see equation~\eqref{Yfinal}.
A proposed resolution of the apparent contradiction with integrability in the S-matrix construction, is to consider the scattering of soliton excitations \cite{Hollowood:2010dt}, which is naturally formulated in the RSOS picture \cite{Hollowood:2010rv}.
This is an additional indication that dealing with the asymptotic behaviour at infinity may shed further light on the classical integrability of these models.

Finally, let us note that SSSG theories were originally defined at the level of their field equations.
They were obtained by applying the Pohlmeyer reduction \cite{Pohlmeyer:1975nb,DAuria:1979ham,DAuria:1980iyh} to symmetric space sigma models.
Subsequently, their Lagrangian was constructed in \cite{Bakas:1995bm} (see also \cite{Fernandez-Pousa:1996aoa} and \cite{Miramontes:2008wt} for a review).
This reduction has been further extended in \cite{Grigoriev:2007bu,Grigoriev:2008jq} to sigma models describing classical strings on $AdS_N \times S^N$ in conformal gauge.
The resulting theories are called Semi-Symmetric Space Sine-Gordon theories and their S-matrices have been investigated in \cite{Hoare:2011fj,Hollowood:2011fq,Hoare:2011nd,Hoare:2013ysa}.
It is also known that the non-ultralocality of these integrable field theories is mild \cite{Delduc:2012mk,Delduc:2012vq} and it would be interesting to generalise the results obtained in the present article.

\paragraph{Acknowledgments.} M.M. thanks J.M. Maillet for useful discussions.
We would like to thank J.L. Miramontes for comments on a draft.
The work of B.H. was supported by a UKRI Future Leaders Fellowship (grant number MR/T018909/1).

\appendix

\section{Explicit construction of \texorpdfstring{$\alpha$}{alpha} and \texorpdfstring{$\gamma$}{gamma} for the spheres}\label{spheres}

In this appendix we give an explicit construction of $\alpha$ and $\gamma$ for $F=SO(N+1)$, $G=SO(N)$, $H=SO(N-1)$.
We use the $(N+1)\times(N+1)$ matrix representation of the Lie algebra $\mathfrak{f} = \mathfrak{so}(N+1)$ realised as
\begin{equation}
\mathcal{G}\alga^T \mathcal{G} = - \alga , \qquad
\alga^\dagger = -\alga ,
\end{equation}
where $\mathcal{G} = \antidiag(1,1,\dots,1)$.
Starting with $\alga \in \mathfrak{gl}(N+1;\mathbb{C})$ the first constraint restricts $\alga$ to lie in $\mathfrak{so}(N+1;\mathbb{C})$, while the second restricts to the real form $\mathfrak{u}(N+1)$.
The intersection of these two Lie algebras is the required real form $\mathfrak{so}(N+1)$.

Denoting the matrix with a 1 in row $i$ and column $j$ and 0 elsewhere by $\mathcal{E}_{i,j}$, for this realisation we can take the Cartan generators of the Cartan-Weyl basis to lie along the diagonal
\begin{equation}\label{eq:cwb1}
h_i = \mathcal{E}_{i,i} - \mathcal{E}_{N+2-i,N+2-i} , \qquad i = 1,\dots,\lfloor \frac{N+1}{2} \rfloor ,
\end{equation}
the positive roots to be upper triangular
\begin{equation}\begin{split}\label{eq:cwb2}
e_{i,j} &= \mathcal{E}_{i,j} - \mathcal{E}_{N+2-j,N+2-i} , \qquad i = 1,\dots,\lfloor \frac{N}{2}\rfloor , \qquad j = i+1,\dots N+1-i ,
\end{split}\end{equation}
and the negative roots to be lower triangular
\begin{equation}\label{eq:cwb3}
f_{i,j} = e_{i,j}^T .
\end{equation}
As usual the generators of the real form, i.e. satisfying $\alga^\dagger = -\alga$, are $i h_i$, $e_{i,j} - f_{i,j}$ and $i(e_{i,j} + f_{i,j})$.

To define the subalgebras $\alg{g} = \mathfrak{so}(N)$ and $\alg{h} = \mathfrak{so}(N-1)$ we introduce the following projection operators
\begin{equation}
\begin{split}
\mathcal{P}_i \alga & = \Ad_{\mathcal{R}}^{-1}\big((\mathcal{I}_{N+1} - \mathcal{E}_{i,i})(\Ad_\mathcal{R}\alga)(\mathcal{I}_{N+1} - \mathcal{E}_{i,i})\big) ,
\\
\mathcal{R} & =
\begin{cases} \operatorname{diag}(\mathcal{I}_{\frac{N-1}{2}},\begin{pmatrix} \frac{1}{\sqrt{2}} & \frac{1}{\sqrt{2}} \\ - \frac{1}{\sqrt{2}} & \frac{1}{\sqrt{2}}\end{pmatrix},\mathcal{I}_{\frac{N-1}{2}}) , \qquad \text{$N$ odd},
\\ \operatorname{diag}(\mathcal{I}_{\frac{N-2}{2}},\begin{pmatrix} \frac{1}{\sqrt{2}} & 0 & \frac{1}{\sqrt{2}} \\ 0 & 1 & 0 \\ - \frac{1}{\sqrt{2}} & 0 & \frac{1}{\sqrt{2}}\end{pmatrix},\mathcal{I}_{\frac{N-2}{2}}) ,\qquad \text{$N$ even},
\end{cases}
\end{split}
\end{equation}
where $\mathcal{I}_n$ is the $n \times n$ identity matrix.
The projectors onto $\alg{g} = \mathfrak{so}(N)$, $\alg{h} = \mathfrak{so}(N-1)$ and $\alg{a} = \mathfrak{so}(2)$ are defined as
\begin{equation}\begin{split}
P^{\mathfrak{g}} & = \mathcal{P}_{\lfloor \frac{N+2}{2} \rfloor}, \qquad
P^{\mathfrak{h}} = \mathcal{P}_{\lfloor \frac{N+4}{2} \rfloor} \mathcal{P}_{\lfloor \frac{N+2}{2} \rfloor}, \qquad
P^{\mathfrak{a}} = 2 \mathcal{I}_{N+1}-\mathcal{P}_{\lfloor \frac{N+4}{2} \rfloor} - \mathcal{P}_{\lfloor \frac{N+2}{2} \rfloor}.
\end{split}
\end{equation}
The generator $\Omega$ of the abelian algebra $\mathfrak{a}$, normalised such that the eigenvalues of $\ad_\Omega$ on $\mathfrak{f}^{\perp\mathbb{C}}$ are $\pm i$, is given by
\begin{equation}
\Omega = \begin{cases}
i h_{\frac{N+1}{2}}, \qquad \text{$N$ odd}
\\
\frac{i}{\sqrt{2}} (e_{\frac{N}{2},\frac{N+2}{2}} + f_{\frac{N}{2},\frac{N+2}{2}}), \qquad \text{$N$ even}
\end{cases}
\end{equation}
A key property of this choice is that the projections of the Cartan-Weyl basis of $\mathfrak{f} = \mathfrak{so}(N+1)$~\eqref{eq:cwb1}--\eqref{eq:cwb3} give Cartan-Weyl bases of $\mathfrak{g} = \mathfrak{so}(N)$ and $\mathfrak{h} = \mathfrak{so}(N-1)$.

The split Drinfel'd-Jimbo R-matrix on $\mathfrak{f}^{\mathbb{C}} = \mathfrak{so}(N+1;\mathbb{C})$ is given by
\begin{equation}
R \alga = \frac{1}{2}\sum_{i,j} \big(\Tr(\alga f_{i,j}) e_{i,j} - \Tr(\alga e_{i,j}) f_{i,j}\big) .
\end{equation}
As is well-known, this operator does not preserve the real form $\alg{f} = \alg{so}(N+1)$.
Nevertheless, for the choice of Cartan-Weyl basis of $\alg{f}$ and subalgebras $\alg{g}$ and $\alg{h}$ above, this R-matrix has the property that $R: \alg{g}^{\mathbb{C}} \to \alg{g}^{\mathbb{C}}$ and $R: \alg{h}^{\mathbb{C}} \to \alg{h}^{\mathbb{C}}$.
Therefore, we can define
\begin{equation}
\alpha = \xi R P^\alg{g},
\end{equation}
which satisfies the split mCYBE on $\alg{g}^{\mathbb{C}}$ and $\alg{f}^{(1)\mathbb{C}} \in \Ker \alpha$ as required.
Moreover, writing
\begin{equation}
\alpha = \alpha^{\alg{h}} + \psi, \qquad \alpha^{\alg{h}} = \alpha P^{\alg{h}}, \qquad \psi = \alpha (P^{\alg{g}} - P^{\alg{h}}),
\end{equation}
we have that
\begin{equation}
\alpha^{\alg{h}} : \alg{h} \to \alg{h}, \qquad P^{\alg{h}}\psi = 0 , \qquad
\psi P^{\alg{h}} = 0 ,
\end{equation}
in agreement with the assumptions outlined in subsection \ref{sec:ccal}.

We now turn to the construction of $\gamma$ that commutes with $\Omega$ and satisfies equations \eqref{cond gamma Ch}, \eqref{cond gamma Om} and \eqref{cond gamma alpha}.
In terms of the corresponding operators
these equations take the form
\begin{equation}
\Ad_\gamma^{-1} P^\alg{h} \Ad_\gamma = P^\alg{h}, \qquad
\Ad_\gamma^{-1} P^\alg{a} \Ad_\gamma = P^\alg{a}, \qquad
\Ad_\gamma^{-1} \alpha^\alg{h} \Ad_\gamma = - \alpha^\alg{h},
\end{equation}
One choice of $\gamma$ that achieves this is
\begin{equation}
\gamma = \mathcal{G}
-\mathcal{E}_{\lfloor\frac{N+1}{2}\rfloor,\lfloor\frac{N+4}{2}\rfloor}
-\mathcal{E}_{\lfloor\frac{N+4}{2}\rfloor,\lfloor\frac{N+1}{2}\rfloor}
+\mathcal{E}_{\lfloor\frac{N+1}{2}\rfloor,\lfloor\frac{N+1}{2}\rfloor}
+\mathcal{E}_{\lfloor\frac{N+4}{2}\rfloor,\lfloor\frac{N+4}{2}\rfloor}.
\end{equation}
Note that this choice is engineered such that $\Ad_\gamma^{-1}$ acts as the identity on $\alg{a}$, as the Cartan involution on $\alg{h}$
\begin{equation}
\Ad_\gamma^{-1} P^{\alg{h}} h_{i} = - P^{\alg{h}} h_i ,
\qquad
\Ad_\gamma^{-1} P^{\alg{h}} e_{i,j} = - P^{\alg{h}} f_{i,j} ,
\qquad
\Ad_\gamma^{-1} P^{\alg{h}} f_{i,j} = - P^{\alg{h}} e_{i,j} ,
\end{equation}
and it maps $\Ad_\gamma^{-1} \omega_\pm \to \omega_\pm$.
Since
\begin{equation}
\mathcal{G}\gamma^T\mathcal{G}\gamma = \mathcal{I}_{N+1}, \qquad \gamma^\dagger \gamma = \mathcal{I}_{N+1}, \qquad
\det \gamma = (-1)^{N}.
\end{equation}
it follows that $\gamma \in O(N+1)$ for even $N$ and $\gamma \in SO(N+1)$ for odd $N$.

\section{Equations satisfied by \texorpdfstring{$r$}{r}, \texorpdfstring{$a$}{a}, \texorpdfstring{$b$}{b}, \texorpdfstring{$c$}{c} and \texorpdfstring{$d$}{d}}
\label{sss context alleviating}

In this appendix we prove the quadratic equations satisfied by the matrices $r$, $a$, $b$, $c$ and $d$.
These equations are used in section \ref{sec4}.

\subsection{Equation satisfied by \texorpdfstring{$r$}{r}}

The matrix $r$ is defined in equation \eqref{defofr12}.
To simplify the presentation, let us define
\beqz
f^0_{12} = \frac{1}{4K} \frac{\lambda_2^2 + \lambda_1^2}{\lambda_2^2 - \lambda_1^2} \qquad
\mbox{and} \qquad f^1_{12} =
\frac{1}{4K} \frac{2 \lambda_1 \lambda_2}{\lambda_2^2 - \lambda_1^2},
\eeqz
where the subscripts indicate the dependence on the spectral parameters.
Using this notation, the matrix $r$ is given by
\beqz
r\ti{12}(\lambda_1,\lambda_2) = f^0_{12} C\ti{12}^{(00)} + f^1_{12}C\ti{12}^{(11)} ,
\eeqz
and
\begin{equation}\begin{aligned} \label{ra ra jac}
[r\ti{12}(\lambda_1,\lambda_2), r\ti{13}(\lambda_1,\lambda_3)]
= & f^0_{12} f^0_{13}[C\ti{12}^{(00)}, C\ti{13}^{(00)}]
+ f^1_{12} f^1_{13}[C\ti{12}^{(11)}, C\ti{13}^{(11)}] \\
& + f^0_{12} f^1_{13}[C\ti{12}^{(00)}, C\ti{13}^{(11)}]
+ f^1_{12} f^0_{13}[C\ti{12}^{(11)}, C\ti{13}^{(00)}].
\end{aligned}\end{equation}
To derive~\eqref{CYBE type ra text}, we sum this commutator with its cyclic permutations (using the fact that $r\ti{12}(\lambda_1,\lambda_2) = - r\ti{21}(\lambda_2,\lambda_1)$).
The sum of first term on the r.h.s. of \eqref{ra ra jac} with its cyclic permutations is the only contribution belonging to $\g\otimes\g\otimes\g$, and is equal to
\beqz
\Bigl(
f^0_{12} f^0_{13} - f^0_{12} f^0_{23} +f^0_{13} f^0_{23}
\Bigr)
[C\ti{12}^{(00)}, C\ti{13}^{(00)}] = \frac{1}{16 K^2}[C\ti{12}^{(00)}, C\ti{13}^{(00)}].
\eeqz
where we use the property \eqref{eq:casimirprop} satisfied by the quadratic Casimir of $\g$.

The remaining terms belong to a tensor product of one copy of $\g$ with two copies of $\f^{(1)}$.
Collecting the terms for which the unique copy of $\g$ is in the first tensorial space, the other two sets of terms can be straightforwardly computed by permutation.
This means that we keep the second term and the relevant cyclic permutations of the third and fourth terms on the r.h.s. of \eqref{ra ra jac}
\begin{equation}\begin{aligned}
f^1_{12} f^1_{13}[C\ti{12}^{(11)}, C\ti{13}^{(11)}] +
&
f^0_{31} f^1_{32}[C\ti{31}^{(00)}, C\ti{32}^{(11)}] +
f^1_{23} f^0_{21}[C\ti{23}^{(11)}, C\ti{21}^{(00)}]
\\ & \qquad
=
\Bigl(f^1_{12} f^1_{13} +
f^0_{31} f^1_{32} +f^1_{23} f^0_{21}\Bigr)
[C\ti{12}^{(11)}, C\ti{13}^{(11)}] = 0,
\end{aligned}\end{equation}
where we have again used standard properties of the quadratic Casimir of $\f$~\eqref{eq:casimirprop}.

Therefore, we conclude that
\begin{equation}\begin{split} \label{CYBE type ra}
[r\ti{12}(\lambda_1,\lambda_2),
r\ti{13}(\lambda_1, \lambda_3)]
+
[r\ti{12}(\lambda_1, \lambda_2),
r\ti{23}(\lambda_2, \lambda_3)]
+
[r\ti{13}(\lambda_1, \lambda_3),
r\ti{23}(\lambda_2, \lambda_3)] \qquad
\\
= \frac{1}{16 K^2}[C^{(00)}\ti{12},C^{(00)}\ti{13}].
\end{split}\end{equation}
The matrix $r$ is not a solution of the CYBE.
However, when $\g$ is abelian, the r.h.s. vanishes and $r$ becomes a solution of the CYBE.
This is the reason why, among the SSSG theories, the Sine-Gordon theory is special.

\subsection{Equations satisfied by \texorpdfstring{$a$}{a} and \texorpdfstring{$d$}{d}}

We recall that $a$ is defined as $a\ti{12}(\lambda_1,\lambda_2) =- r\ti{12}(\lambda_1,\lambda_2) - \alpha\ti{12}$ where $\alpha \in \alg{g} \otimes \alg{g}$ is skew-symmetric, independent of the spectral parameter and solves
\beq \label{alpha mcybe appendix}
[\alpha\ti{12},\alpha\ti{13}]
+
[\alpha\ti{12},\alpha\ti{23}]
+[\alpha\ti{13},\alpha\ti{23}] = -\xi^2 [C^{(00)}\ti{12},C^{(00)}\ti{13}].
\eeq
In the expansion of
\beq \label{dev of rpm rpm}
[a \ti{12}(\lambda_1,\lambda_2),
a\ti{13}(\lambda_1, \lambda_3)]
+
[a\ti{12}(\lambda_1, \lambda_2),
a\ti{23}(\lambda_2, \lambda_3)]
+
[a\ti{13}(\lambda_1, \lambda_3),
a\ti{23}(\lambda_2, \lambda_3)],
\eeq
the contributions coming from the terms quadratic in $r$ or quadratic in $\alpha$ follow immediately from equations \eqref{CYBE type ra} and \eqref{alpha mcybe appendix}.
To analyse the mixed terms, let us consider the contribution from the first term in \eqref{dev of rpm rpm}
\begin{equation}\begin{aligned} \label{eq int a alpha}
& [f^0_{12} C\ti{12}^{(00)} + f^1_{12} C\ti{12}^{(11)}, \alpha\ti{13}]
+[\alpha\ti{12}, f^0_{13} C\ti{13}^{(00)} + f^1_{13} C\ti{13}^{(11)}]
\\
& \quad = [f^0_{12} C\ti{12}^{(00)} + f^1_{12} C\ti{12}^{(11)}, \alpha\ti{13}]
-[f^0_{13} C\ti{13}^{(00)} + f^1_{13} C\ti{13}^{(11)},\alpha\ti{23}],
\end{aligned}\end{equation}
where we have used that $\alpha \in \g \otimes \g$ is skew-symmetric, together with the identities~\eqref{eq:casimirprop}.
Summing~\eqref{eq int a alpha} with the cyclic permutations, we find that the six terms cancel amongst each other.
It follows that $a$ is solution of
\begin{equation*}\begin{split}
[a \ti{12}(\lambda_1,\lambda_2),
a\ti{13}(\lambda_1, \lambda_3)]
+
[a\ti{12}(\lambda_1, \lambda_2),
a\ti{23}(\lambda_2, \lambda_3)]
+
[a\ti{13}(\lambda_1, \lambda_3),
a\ti{23}(\lambda_2, \lambda_3)] \qquad
\\ = (\frac{1}{16 K^2}-\xi^2)[C\ti{12}^{(00)},C\ti{13}^{(00)}].
\end{split}\end{equation*}
Similarly, the matrix $d=- r +\alpha$ is solution of the same equation.
Furthermore, when $\xi^2=1/(16 K^2)$, $a$ and $d$ become solutions of the CYBE.

\subsection{Equations satisfied by \texorpdfstring{$a$}{a} with \texorpdfstring{$c$}{c} and \texorpdfstring{$d$}{d} with \texorpdfstring{$b$}{b}}

We recall that $c=s-\alpha$ with $s\ti{12}=\frac{1}{4K}C\ti{12}^{(00)}$.
The matrix $c$ is therefore independent of the spectral parameter and valued in $\alg{g} \otimes \alg{g}$.
Expanding out $a$ and $c$, we have
\begin{equation}\begin{split}\label{acex}
& [a\ti{12}(\lambda_1,\lambda_2),c\ti{13}] +
[a\ti{12}(\lambda_1,\lambda_2),c\ti{23}]+
[c\ti{13},c\ti{23}]
\\
& = - [r\ti{12}(\lambda_1,\lambda_2), c\ti{13}+c\ti{23}]
+[s\ti{13}, \alpha\ti{12} + \alpha\ti{32}]
+[s\ti{23}, \alpha\ti{12} + \alpha\ti{13}]
\\
& \qquad + [\alpha\ti{12},\alpha\ti{13}] + [\alpha\ti{12},\alpha\ti{23}] + [\alpha\ti{13},\alpha\ti{23}]
+ [s\ti{13},s\ti{23}],
\end{split}\end{equation}
where we have used the antisymmetry of $\alpha$.
Since both $c$ and $\alpha$ belong to $\g\otimes\g$, and $r\ti{12}(\lambda_1,\lambda_2)$ and $s\ti{12}$ are linear combinations of $C\ti{12}^{(00)}$ and $C\ti{12}^{(11)}$, it immediately follows from the identities \eqref{eq:casimirprop} that the first line of the r.h.s. of~\eqref{acex} vanishes.
Substituting in for $s$ in the second line and using~\eqref{alpha mcybe appendix}, we conclude that
\beqz
[a\ti{12},c\ti{13}] +[a\ti{12},c\ti{23}]+[c\ti{13},c\ti{23}]
=(\frac{1}{16 K^2}-\xi^2) [C\ti{12}^{(00)},C\ti{13}^{(00)}].
\eeqz
The identity
\beqz
[d\ti{12},b\ti{13}] +[d\ti{12},b\ti{23}]+[b\ti{13},b\ti{23}],
=(\frac{1}{16 K^2}-\xi^2) [C\ti{12}^{(00)},C\ti{13}^{(00)}],
\eeqz
similarly follows by sending $\alpha$ to $-\alpha$.

\section{Derivation of the Poisson bracket of transition matrices on the lattice}\label{ttpbproof}

\def\L{{\cal L}}
In this appendix we derive the Poisson bracket of transition matrices on the lattice in equation~\eqref{eq:ttpb}.
In the following we will suppress all dependence on the spectral parameters.
For $n \geq q$ we have
\begin{equation}
T^{n,q} = \L^n \L^{n-1} \dots \L^{q+1} \L^q ~.
\end{equation}
For convenience, we also define $T^{n,q} = \id$ for $n<q$.

Using the Leibniz rule, we have
\begin{equation}
\{T\ti{1}^{n,q},T\ti{2}^{m,p}\}
= \sum_{r=0}^{n-q}\sum_{s=0}^{m-p} T\ti{1}^{n,q+r+1}T\ti{2}^{m,p+s+1}\{\L\ti{1}^{q+r},\L\ti{2}^{p+s}\}
T\ti{1}^{q+r-1,q}T\ti{2}^{p+s-1,p}.
\end{equation}
Substituting in the Poisson bracket for $\L$ in equation~\eqref{lattice algebra 2} then gives
\begin{equation}\begin{split}
\{T\ti{1}^{n,q},T\ti{2}^{m,p}\}
= \sum_{r=0}^{n-q}\sum_{s=0}^{m-p}
\Big(
& T\ti{1}^{n,q+r+1}T\ti{2}^{m,p+s+1} a\ti{12} T\ti{1}^{q+r,q}T\ti{2}^{p+s,p} \delta_{q+r+1,p+s+1}
\\ & -
T\ti{1}^{n,q+r}T\ti{2}^{m,p+s} d\ti{12} T\ti{1}^{q+r-1,q}T\ti{2}^{p+s-1,p} \delta_{q+r,p+s}
\\ & +
T\ti{1}^{n,q+r}T\ti{2}^{m,p+s+1} b\ti{12} T\ti{1}^{q+r-1,q}T\ti{2}^{p+s,p} \delta_{q+r,p+s+1}
\\ & -
T\ti{1}^{n,q+r+1}T\ti{2}^{m,p+s} c\ti{12} T\ti{1}^{q+r,q}T\ti{2}^{p+s-1,p} \delta_{q+r+1,p+s}
\Big),
\end{split}\end{equation}
where we have used $\delta_{q+r,p+s} = \delta_{q+r+1,p+s+1}$ in the term proportional to $a\ti{12}$.
We now shift the summed indices so that the Kronecker deltas in all four terms are the same
\begin{equation}\begin{split}
\{T\ti{1}^{n,q},T\ti{2}^{m,p}\}
=
&\sum_{r=1}^{n-q+1}\sum_{s=1}^{m-p+1}
T\ti{1}^{n,q+r}T\ti{2}^{m,p+s} a\ti{12} T\ti{1}^{q+r-1,q}T\ti{2}^{p+s-1,p} \delta_{q+r,p+s}
\\& -
\sum_{r=0}^{n-q}\sum_{s=0}^{m-p}
T\ti{1}^{n,q+r}T\ti{2}^{m,p+s} d\ti{12} T\ti{1}^{q+r-1,q}T\ti{2}^{p+s-1,p} \delta_{q+r,p+s}
\\ & +
\sum_{r=0}^{n-q}\sum_{s=1}^{m-p+1}
T\ti{1}^{n,q+r}T\ti{2}^{m,p+s} b\ti{12} T\ti{1}^{q+r-1,q}T\ti{2}^{p+s-1,p} \delta_{q+r,p+s}
\\ & -
\sum_{r=1}^{n-q+1}\sum_{s=0}^{m-p}
T\ti{1}^{n,q+r}T\ti{2}^{m,p+s} c\ti{12} T\ti{1}^{q+r-1,q}T\ti{2}^{p+s-1,p} \delta_{q+r,p+s}.
\end{split}\end{equation}
Next, we split up the sums in the following way
\begin{equation}\begin{split}
\{T\ti{1}^{n,q},T\ti{2}^{m,p}\}
=
&\sum_{r=1}^{n-q}\sum_{s=1}^{m-p}
T\ti{1}^{n,q+r}T\ti{2}^{m,p+s} (a\ti{12} -d\ti{12} + b\ti{12}-c\ti{12}) T\ti{1}^{q+r-1,q}T\ti{2}^{p+s-1,p} \delta_{q+r,p+s}
\\
& + \sum_{r=1}^{n-q}T\ti{1}^{n,q+r}\big((a\ti{12} + b\ti{12})T\ti{2}^{m,p}\delta_{m+1,q+r}
-
T\ti{2}^{m,p}(d\ti{12} + c\ti{12})\delta_{p,q+r}\big)T\ti{1}^{q+r-1,q}
\\ & +
\sum_{s=1}^{m-p} T\ti{2}^{m,p+s}\big((a\ti{12} - c\ti{12})T\ti{1}^{n,q}\delta_{p+s,n+1}
-
T\ti{1}^{n,q}(d\ti{12} - b\ti{12})\delta_{p+s,q}\big)T\ti{2}^{p+s-1,p}
\\ & +
a\ti{12} T\ti{1}^{n,q} T\ti{2}^{m,p} \delta_{m,n} - T\ti{1}^{n,q} T\ti{2}^{m,p} d\ti{12} \delta_{p,q}
+ T\ti{1}^{n,q} b\ti{12} T\ti{2}^{m,p} \delta_{m+1,q} - T\ti{2}^{m,p} c\ti{12} T\ti{1}^{n,q} \delta_{p,n+1} .
\end{split}\end{equation}
Finally, we use the relation~\eqref{spec rela abcd} to notice that the first line of the r.h.s. vanishes.
We are then left with~\eqref{eq:ttpb} as claimed.

\bibliographystyle{JHEP}
\bibliography{fr,fr2,fr3,fr4,fr5}

\providecommand{\href}[2]{#2}\begingroup\raggedright\begin{thebibliography}{10}

\bibitem{Hollowood:2009tw}
T.~J. Hollowood and J.~L. Miramontes, \emph{{Magnons, their Solitonic Avatars
  and the Pohlmeyer Reduction}},
  \href{http://dx.doi.org/10.1088/1126-6708/2009/04/060}{\emph{JHEP} {\bf 04}
  (2009) 060}, [\href{https://arxiv.org/abs/0902.2405}{{\tt 0902.2405}}].

\bibitem{Hollowood:2011fm}
T.~J. Hollowood and J.~L. Miramontes, \emph{{The Semi-Classical Spectrum of
  Solitons and Giant Magnons}},
  \href{http://dx.doi.org/10.1007/JHEP05(2011)062}{\emph{JHEP} {\bf 05} (2011)
  062}, [\href{https://arxiv.org/abs/1103.3148}{{\tt 1103.3148}}].

\bibitem{Hollowood:2013oca}
T.~J. Hollowood, J.~Miramontes and D.~M. Schmidtt, \emph{{The Structure of
  Non-Abelian Kinks}},
  \href{http://dx.doi.org/10.1007/JHEP10(2013)058}{\emph{JHEP} {\bf 10} (2013)
  058}, [\href{https://arxiv.org/abs/1306.6651}{{\tt 1306.6651}}].

\bibitem{fadtakbook87}
L.~Faddeev and L.~L.A.~Takhtajan, \emph{{Hamiltonian methods in the theory of
  solitons}}.
\newblock Springer, 1987.

\bibitem{Babebook}
O.~Babelon, D.~Bernard and M.~Talon, \emph{{Introduction to Classical
  Integrable Models}}.
\newblock Cambridge University Press, 2003.

\bibitem{Faddeev:1979gh}
L.~Faddeev, E.~Sklyanin and L.~Takhtajan, \emph{{The Quantum Inverse Problem
  Method. 1}}, {\emph{Theor. Math. Phys.} {\bf 40} (1980) 688--706}.

\bibitem{Sklyanin:1980ij}
E.~Sklyanin, \emph{{Quantum version of the method of inverse scattering
  problem}}, \href{http://dx.doi.org/10.1007/BF01091462}{\emph{J. Sov. Math.}
  {\bf 19} (1982) 1546--1596}.

\bibitem{Hollowood:2010dt}
T.~J. Hollowood and J.~L. Miramontes, \emph{{Classical and Quantum Solitons in
  the Symmetric Space Sine-Gordon Theories}},
  \href{http://dx.doi.org/10.1007/JHEP04(2011)119}{\emph{JHEP} {\bf 1104}
  (2011) 119}, [\href{https://arxiv.org/abs/1012.0716}{{\tt 1012.0716}}].

\bibitem{Maillet:1985fn}
J.~M. Maillet, \emph{{Kac-Moody algebra and extended Yang-Baxter relations in
  the $O(N)$ non-linear sigma model}},
  \href{http://dx.doi.org/10.1016/0370-2693(85)91075-5}{\emph{Phys. Lett.} {\bf
  B162} (1985) 137}.

\bibitem{Maillet:1985ek}
J.~M. Maillet, \emph{{New integrable canonical structures in two-dimensional
  models}}, \href{http://dx.doi.org/10.1016/0550-3213(86)90365-2}{\emph{Nucl.
  Phys.} {\bf B269} (1986) 54}.

\bibitem{Delduc:2012qb}
F.~Delduc, M.~Magro and B.~Vicedo, \emph{{Alleviating the non-ultralocality of
  coset sigma models through a generalized Faddeev-Reshetikhin procedure}},
  \href{http://dx.doi.org/10.1007/JHEP08(2012)019}{\emph{JHEP} {\bf 1208}
  (2012) 019}, [\href{https://arxiv.org/abs/1204.0766}{{\tt 1204.0766}}].

\bibitem{Delduc:2012mk}
F.~Delduc, M.~Magro and B.~Vicedo, \emph{{A lattice Poisson algebra for the
  Pohlmeyer reduction of the $AdS_5 \times S^5$ superstring}}, {\emph{Phys.
  Lett.} {\bf B713} (2012) 347--349},
  [\href{https://arxiv.org/abs/1204.2531}{{\tt 1204.2531}}].

\bibitem{Freidel:1991jx}
L.~Freidel and J.~M. Maillet, \emph{{Quadratic algebras and integrable
  systems}}, \href{http://dx.doi.org/10.1016/0370-2693(91)91566-E}{\emph{Phys.
  Lett.} {\bf B262} (1991) 278--284}.

\bibitem{Freidel:1991jv}
L.~Freidel and J.~M. Maillet, \emph{{On classical and quantum integrable field
  theories associated to Kac-Moody current algebras}},
  \href{http://dx.doi.org/10.1016/0370-2693(91)90479-A}{\emph{Phys. Lett.} {\bf
  B263} (1991) 403--410}.

\bibitem{Vicedo:2017cge}
B.~Vicedo, \emph{{On integrable field theories as dihedral affine Gaudin
  models}}, \href{http://dx.doi.org/10.1093/imrn/rny128}{\emph{International
  Mathematics Research Notices} {\bf rny128} (2017) },
  [\href{https://arxiv.org/abs/1701.04856}{{\tt 1701.04856}}].

\bibitem{Pohlmeyer:1975nb}
K.~Pohlmeyer, \emph{{Integrable hamiltonian systems and interactions through
  quadratic constraints}},
  \href{http://dx.doi.org/10.1007/BF01609119}{\emph{Commun. Math. Phys.} {\bf
  46} (1976) 207--221}.

\bibitem{Lund:1976ze}
F.~Lund and T.~Regge, \emph{{Unified approach to strings and vortices with
  soliton solutions}},
  \href{http://dx.doi.org/10.1103/PhysRevD.14.1524}{\emph{Phys. Rev.} {\bf D14}
  (1976) 1524}.

\bibitem{Lund:1977dt}
F.~Lund, \emph{{Example of a relativistic, completely integrable, hamiltonian
  system}}, \href{http://dx.doi.org/10.1103/PhysRevLett.38.1175}{\emph{Phys.
  Rev. Lett.} {\bf 38} (1977) 1175}.

\bibitem{Getmanov:1977hk}
B.~Getmanov, \emph{{New Lorentz invariant systems with exact multi-soliton
  solutions}}, {\emph{JETP Lett.} {\bf 25} (1977) 119}.

\bibitem{DAuria:1979ham}
R.~D'Auria, T.~Regge and S.~Sciuto, \emph{{A general scheme for bidimensional
  models with associate linear set}},
  \href{http://dx.doi.org/10.1016/0370-2693(80)90143-4}{\emph{Phys. Lett.} {\bf
  B89} (1980) 363}.

\bibitem{DAuria:1980iyh}
R.~D'Auria, T.~Regge and S.~Sciuto, \emph{{Group Theoretical Construction of
  Two-dimensional Models With Infinite Set of Conservation Laws}},
  \href{http://dx.doi.org/10.1016/0550-3213(80)90366-1}{\emph{Nucl. Phys. B}
  {\bf 171} (1980) 167--188}.

\bibitem{Bakas:1995bm}
I.~Bakas, Q.-H. Park and H.-J. Shin, \emph{{Lagrangian formulation of symmetric
  space sine-Gordon models}},
  \href{http://dx.doi.org/10.1016/0370-2693(96)00026-3}{\emph{Phys. Lett.} {\bf
  B372} (1996) 45--52}, [\href{https://arxiv.org/abs/hep-th/9512030}{{\tt
  hep-th/9512030}}].

\bibitem{Fernandez-Pousa:1996aoa}
C.~R. Fernandez-Pousa, M.~V. Gallas, T.~J. Hollowood and J.~Miramontes,
  \emph{{The Symmetric space and homogeneous sine-Gordon theories}},
  \href{http://dx.doi.org/10.1016/S0550-3213(96)00603-7}{\emph{Nucl. Phys.}
  {\bf B484} (1997) 609--630},
  [\href{https://arxiv.org/abs/hep-th/9606032}{{\tt hep-th/9606032}}].

\bibitem{Miramontes:2008wt}
J.~L. Miramontes, \emph{{Pohlmeyer reduction revisited}},
  \href{http://dx.doi.org/10.1088/1126-6708/2008/10/087}{\emph{JHEP} {\bf 10}
  (2008) 087}, [\href{https://arxiv.org/abs/0808.3365}{{\tt 0808.3365}}].

\bibitem{Hollowood:1994vx}
T.~J. Hollowood, J.~Miramontes and Q.-H. Park, \emph{{Massive integrable
  soliton theories}},
  \href{http://dx.doi.org/10.1016/0550-3213(95)00142-F}{\emph{Nucl.Phys.} {\bf
  B445} (1995) 451--468}, [\href{https://arxiv.org/abs/hep-th/9412062}{{\tt
  hep-th/9412062}}].

\bibitem{Alekseev:1998mc}
A.~{\relax Yu}. Alekseev and V.~Schomerus, \emph{{D-branes in the WZW model}},
  \href{http://dx.doi.org/10.1103/PhysRevD.60.061901}{\emph{Phys. Rev.} {\bf
  D60} (1999) 061901}, [\href{https://arxiv.org/abs/hep-th/9812193}{{\tt
  hep-th/9812193}}].

\bibitem{Gawedzki:2001rm}
K.~Gawedzki, I.~Todorov and P.~Tran-Ngoc-Bich, \emph{{Canonical quantization of
  the boundary Wess-Zumino-Witten model}},
  \href{http://dx.doi.org/10.1007/s00220-004-1107-6}{\emph{Commun. Math. Phys.}
  {\bf 248} (2004) 217--254}, [\href{https://arxiv.org/abs/hep-th/0101170}{{\tt
  hep-th/0101170}}].

\bibitem{Gawedzki:2001ye}
K.~Gawedzki, \emph{{Boundary WZW, $G/H$, $G/G$ and CS theories}},
  \href{http://dx.doi.org/10.1007/s00023-002-8639-0}{\emph{Annales Henri
  Poincare} {\bf 3} (2002) 847--881},
  [\href{https://arxiv.org/abs/hep-th/0108044}{{\tt hep-th/0108044}}].

\bibitem{Elitzur:2001qd}
S.~Elitzur and G.~Sarkissian, \emph{{D branes on a gauged WZW model}},
  \href{http://dx.doi.org/10.1016/S0550-3213(02)00010-X}{\emph{Nucl. Phys. B}
  {\bf 625} (2002) 166--178}, [\href{https://arxiv.org/abs/hep-th/0108142}{{\tt
  hep-th/0108142}}].

\bibitem{Figueroa-OFarrill:2005vws}
J.~M. Figueroa-O'Farrill and N.~Mohammedi, \emph{{Gauging the Wess-Zumino term
  of a sigma model with boundary}},
  \href{http://dx.doi.org/10.1088/1126-6708/2005/08/086}{\emph{JHEP} {\bf 08}
  (2005) 086}, [\href{https://arxiv.org/abs/hep-th/0506049}{{\tt
  hep-th/0506049}}].

\bibitem{Delduc:2019bcl}
F.~Delduc, S.~Lacroix, M.~Magro and B.~Vicedo, \emph{{Assembling integrable
  $\sigma$-models as affine Gaudin models}},
  \href{http://dx.doi.org/10.1007/JHEP06(2019)017}{\emph{JHEP} {\bf 06} (2019)
  017}, [\href{https://arxiv.org/abs/1903.00368}{{\tt 1903.00368}}].

\bibitem{SemenovTianShansky:1994dm}
M.~Semenov-Tian-Shansky, \emph{{Monodromy map and classical $R$-matrices}},
  {\emph{Journal of Math. Sciences} {\bf 77} (1995) 3226},
  [\href{https://arxiv.org/abs/hep-th/9402054}{{\tt hep-th/9402054}}].

\bibitem{SemenovTianShansky:1995ha}
M.~Semenov-Tian-Shansky and A.~Sevostyanov, \emph{{Classical and quantum
  nonultralocal systems on the lattice}},
  \href{https://arxiv.org/abs/hep-th/9509029}{{\tt hep-th/9509029}}.

\bibitem{Delduc:2015xdm}
F.~Delduc, S.~Lacroix, M.~Magro and B.~Vicedo, \emph{{On the Hamiltonian
  integrability of the bi-Yang-Baxter $\sigma$-model}},
  \href{http://dx.doi.org/10.1007/JHEP03(2016)104}{\emph{JHEP} {\bf 03} (2016)
  104}, [\href{https://arxiv.org/abs/1512.02462}{{\tt 1512.02462}}].

\bibitem{Vicedo:2010qd}
B.~Vicedo, \emph{{The classical R-matrix of AdS/CFT and its Lie dialgebra
  structure}}, \href{http://dx.doi.org/10.1007/s11005-010-0446-9}{\emph{Lett.
  Math. Phys.} {\bf 95} (2011) 249--274},
  [\href{https://arxiv.org/abs/1003.1192}{{\tt 1003.1192}}].

\bibitem{Adler:1979ib}
M.~Adler, \emph{{On a Trace functional for formal pseudo differential operators
  and the symplectic structure of the Korteweg-de Vries equation}},
  \href{http://dx.doi.org/10.1007/BF01410079}{\emph{Invent. Math.} {\bf 50}
  (1979) 219--248}.

\bibitem{Kostant:1979qu}
B.~Kostant, \emph{{The Solution to a generalized Toda lattice and
  representation theory}},
  \href{http://dx.doi.org/10.1016/0001-8708(79)90057-4}{\emph{Adv. Math.} {\bf
  34} (1979) 195--338}.

\bibitem{Symes:1981}
W.~W. Symes, \emph{{Systems of Toda type, inverse spectral problems, and
  representation theory}},
  \href{http://dx.doi.org/10.1007/BF01389068}{\emph{Inventiones mathematicae}
  {\bf 63} (1981) 519--519}.

\bibitem{SemenovTianShansky:1983ik}
M.~Semenov-Tian-Shansky, \emph{{What is a classical $r$-matrix?}},
  {\emph{Funct. Anal. Appl.} {\bf 17} (1983) 259--272}.

\bibitem{erlove77}
E.~R. Love, \emph{{Repeated Singular Integrals}},
  \href{http://dx.doi.org/10.1112/jlms/s2-15.1.99}{\emph{Journal of the London
  Mathematical Society} {\bf s2-15} (01, 1977) 99--102}.

\bibitem{Hoare:2010fb}
B.~Hoare and A.~Tseytlin, \emph{{On the perturbative S-matrix of generalized
  sine-Gordon models}},
  \href{http://dx.doi.org/10.1007/JHEP11(2010)111}{\emph{JHEP} {\bf 1011}
  (2010) 111}, [\href{https://arxiv.org/abs/1008.4914}{{\tt 1008.4914}}].

\bibitem{Hollowood:2010rv}
T.~J. Hollowood and J.~L. Miramontes, \emph{{The Relativistic Avatars of Giant
  Magnons and their S-Matrix}},
  \href{http://dx.doi.org/10.1007/JHEP10(2010)012}{\emph{JHEP} {\bf 1010}
  (2010) 012}, [\href{https://arxiv.org/abs/1006.3667}{{\tt 1006.3667}}].

\bibitem{Grigoriev:2007bu}
M.~Grigoriev and A.~A. Tseytlin, \emph{{Pohlmeyer reduction of AdS$_5$ $\times$
  S$^5$ superstring sigma model}},
  \href{http://dx.doi.org/10.1016/j.nuclphysb.2008.01.006}{\emph{Nucl. Phys.}
  {\bf B800} (2008) 450--501}, [\href{https://arxiv.org/abs/0711.0155}{{\tt
  0711.0155}}].

\bibitem{Grigoriev:2008jq}
M.~Grigoriev and A.~A. Tseytlin, \emph{{On reduced models for superstrings on
  AdS$_n$ $\times$ S$^n$}},
  \href{http://dx.doi.org/10.1142/S0217751X08040652}{\emph{Int. J. Mod. Phys.}
  {\bf A23} (2008) 2107--2117}, [\href{https://arxiv.org/abs/0806.2623}{{\tt
  0806.2623}}].

\bibitem{Hoare:2011fj}
B.~Hoare and A.~Tseytlin, \emph{{Towards the quantum S-matrix of the Pohlmeyer
  reduced version of $AdS_5 \times S^5$ superstring theory}},
  \href{http://dx.doi.org/10.1016/j.nuclphysb.2011.05.016}{\emph{Nucl. Phys.}
  {\bf B851} (2011) 161--237}, [\href{https://arxiv.org/abs/1104.2423}{{\tt
  1104.2423}}].

\bibitem{Hollowood:2011fq}
T.~J. Hollowood and J.~Miramontes, \emph{{The $AdS_5 \times S_5$ Semi-Symmetric
  Space Sine-Gordon Theory}},
  \href{http://dx.doi.org/10.1007/JHEP05(2011)136}{\emph{JHEP} {\bf 1105}
  (2011) 136}, [\href{https://arxiv.org/abs/1104.2429}{{\tt 1104.2429}}].

\bibitem{Hoare:2011nd}
B.~Hoare, T.~J. Hollowood and J.~L. Miramontes, \emph{{A Relativistic Relative
  of the Magnon S-Matrix}},
  \href{http://dx.doi.org/10.1007/JHEP11(2011)048}{\emph{JHEP} {\bf 1111}
  (2011) 048}, [\href{https://arxiv.org/abs/1107.0628}{{\tt 1107.0628}}].

\bibitem{Hoare:2013ysa}
B.~Hoare, T.~J. Hollowood and J.~L. Miramontes, \emph{{Restoring Unitarity in
  the $q$-Deformed World-Sheet S-Matrix}},
  \href{http://dx.doi.org/10.1007/JHEP10(2013)050}{\emph{JHEP} {\bf 1310}
  (2013) 050}, [\href{https://arxiv.org/abs/1303.1447}{{\tt 1303.1447}}].

\bibitem{Delduc:2012vq}
F.~Delduc, M.~Magro and B.~Vicedo, \emph{{Alleviating the non-ultralocality of
  the $AdS_5 \times S^5$ superstring}},
  \href{http://dx.doi.org/10.1007/JHEP10(2012)061}{\emph{JHEP} {\bf 1210}
  (2012) 061}, [\href{https://arxiv.org/abs/1206.6050}{{\tt 1206.6050}}].

\end{thebibliography}\endgroup
\end{document}